\documentclass[final, 5p]{elsarticle}

\usepackage{lineno,hyperref}
\modulolinenumbers[5]

\journal{Materials Characterization}

\hypersetup{
    colorlinks=true,
    linkcolor=blue,
    citecolor=blue,
    filecolor=blue,      
    urlcolor=blue,
}
\newcommand{\vtheta}{\boldsymbol{\theta}}
\newcommand{\vu}{\boldsymbol{u}}
\newcommand{\vv}{\boldsymbol{v}}

\newcommand{\mysize}{1.0}
\newcommand{\mysizehalf}{0.45}
\newcommand{\mysizethird}{0.31}

\usepackage{amsmath, amssymb}
\usepackage{booktabs} %for topline, bottomline
\usepackage{siunitx}
\usepackage{enumitem} % for aviding bullet point indent [leftmargin=5pt] (original 26 pt)
\usepackage[caption=false]{subfig}
\usepackage{hyphenat} %use \hyp{} instead of "-"  for line breaks in hyphened words

\makeatletter
\def\ps@pprintTitle{%
  \let\@oddhead\@empty
  \let\@evenhead\@empty
  \def\@oddfoot{ \footnotesize \parbox[t]{1.2\linewidth}{\textit{Accepted manuscript, Materials Characterization \\ DOI: \href{https://doi.org/10.1016/j.matchar.2020.110796}{10.1016/j.matchar.2020.110796} } \\ \copyright 2020. \textit{Licensed under the  \href{http://creativecommons.org/licenses/by-nc-nd/4.0}{CC-BY-NC-ND 4.0} license.  } } \hfill\textit{ November 27, 2020}} %
  \let\@evenfoot\@oddfoot
 }
\makeatother

\begin{document}

%\linenumbers
\begin{frontmatter}

\title{\large A Physical Model for Microstructural Characterization and Segmentation of 3D Tomography Data}

\author[address_energy]{Elise O. Brenne\corref{mycorrespondingauthor}}
\cortext[mycorrespondingauthor]{Corresponding author}
\ead{elbre@dtu.dk}

\author[address_compute]{Vedrana A. Dahl }
\author[address_energy]{Peter S. J\o rgensen}

\address[address_energy]{Department of Energy Conversion and Storage, Technical University of Denmark, Fysikvej, 2800 Kgs. Lyngby, Denmark}
\address[address_compute]{Department of Applied Mathematics and Computer Science, Technical University of Denmark, Richard Petersens Plads, 2800 Kgs. Lyngby, Denmark}

\begin{abstract}
We present a novel method for characterizing the microstructure of a material from volumetric datasets such as 3D image data from computed tomography (CT). The method is based on a new statistical model for the distribution of voxel intensities and gradient magnitudes, incorporating prior knowledge about the physical nature of the imaging process. It allows for direct quantification of parameters of the imaged sample like volume fractions, interface areas and material density, and parameters related to the imaging process like image resolution and noise levels.

Existing methods for characterization from 3D images often require segmentation of the data, a procedure where each voxel is labeled according to the best guess of which material it represents. Through our approach, the segmentation step is circumvented so that errors and computational costs related to this part of the image processing pipeline are avoided. Instead, the material parameters are quantified through their known relation to parameters of our model which is fitted directly to the raw, unsegmented data.  We present an automated model fitting procedure that gives reproducible results without human bias and enables automatic analysis of large sets of tomograms.

For more complex structure analysis questions, a segmentation is still beneficial. We show that our model can be used as input to existing probabilistic methods, providing a segmentation that is based on the physics of the imaged sample. Because our model accounts for mixed-material voxels stemming from blurring inherent to the imaging technique, we reduce the errors that other methods can create at interfaces between materials.

\end{abstract}

\begin{keyword}
X-ray tomography\sep 3D image analysis\sep microstructural characterization\sep physical parameter extraction\sep automated data analysis\sep Gaussian mixture model
\end{keyword}

\end{frontmatter}

%%%%%%%%%%%%%%%%%%%%%%%%%%%%%%%%%%%%%%%%%%%%%%%%%%%%%%%%%%%%%%%%%%%%%%%%%%%%%%%%%%%%%%%%%%%
\section{Introduction}
%\the\columnwidth
In materials science, 3D tomographic imaging is becoming a powerful tool for investigating complex relations between material microstructure and material properties~\cite{Salvo2010, Maire2014}. The performance of devices like batteries and fuel cells is largely influenced by the microstructure of the constituent materials~\cite{DeAngelis2018}. Accurate measurements of characteristics like volume fractions and interface areas are therefore important to understand and optimize the device. Therefore, the quantification of such material structure parameters presents a challenge that is receiving much attention. 

While the methods presented in this paper are applicable to volumetric data from a range of imaging modalities like neutron, electron or visible light tomography, we here focus on X-ray computed tomography (CT). Typically, the CT workflow starts with acquisition of projection data and 3D tomographic reconstruction. Next, the image data is segmented, i.e. each voxel is labeled according to the material phase it represents. Thereafter, structural parameters can be measured in the segmented data~\cite{Jorgensen2010}, and geometries can be extracted for simulation of physical properties. This way, the quality of the segmentation is directly affecting the accuracy of the estimated material parameters~\cite{Saxena2017}. 

This dependence on the segmentation raises a number of issues. Currently, many segmentation methods rely on visual inspection for parameter tuning, and methods based on supervised or semi-supervised learning require manual labeling of training data~\cite{Enguehard2019}, introducing an operator bias~\cite{Iassonov2009, Hoyte2011}. This may lead to systematic errors and inconsistent, misleading results. Moreover, the visual reliance makes it challenging for others to reproduce the results and to assess the uncertainty in extracted material parameters. In addition, steady increases in hardware and software capabilities are leading to larger datasets. Techniques like \textit{in-situ} tomography generating time-series data is one example where huge amounts of data render a manual assessment of the segmentation quality infeasible. 

One approach to address these problems is to use a statistical model for quantification, circumventing the segmentation step. Instead of measuring sample properties in segmented image data, this approach utilises information in the distribution of voxel intensities, i.e. the data histogram. Through fitting a model to the data which has physically meaningful model parameters, information about the imaged sample can be extracted. One well-known example of such a physical model is the Gaussian mixture model (GMM)~\cite{McLachlan2019}, which models the intensity distribution of each phase in a dataset as a Gaussian. As an example, in a tomogram of a porous structure, the two phases (pores and solid material) will be modelled with two Gaussian components. The mean, variance and weight of each component is then interpreted as the intensity, the noise level and the amount of each phase, respectively~\cite{Dubes1990, Moonen2019}. Therefore, through fitting the GMM to the data, one gets estimates of these physical parameters related to the sample structure and imaging procedure. The main shortcoming of the GMM is that it does not take into account the blurred edges or interfaces between materials resulting from the finite resolution of the imaging pipeline. Gage et al. introduces a model attempting to capture blurring due to the partial volume effect~\cite{Gage1992}, extending the GMM by adding components modelling the interfaces. We refer to this model as the partial volume mixture model (PVMM). This model is reasonable when discretization of the data is the limiting factor for the image resolution, but gives a poor fit to the data when other blurring effects are dominating~\cite{Kindlmann1998, Bentzen1983, Nickoloff1985}. Both the GMM and the PVMM models the image intensities only. They therefore perform less well for data with more than two phases, where the intensity of voxels on the interface between phases may overlap with the intensity of other phase interiors. 

In addition to quantification, statistical models like the GMM and PVMM can be exploited for segmentation through \textit{probabilistic} segmentation methods, e.g. maximum-likelihood segmentation or methods based on Markov Random Fields~\cite{Hassner1981, Berthod1996}. The GMM is often used for simplicity, but the methods can be modified to incorporate other mixture models~\cite{Blekas2005, Sanjay-Gopal1998, Zhang2001, Simmons2009}. For example, a range of methods for segmentation have been developed based on the PVMM~\cite{Laidlaw1998, Tohka2004}. 
One advantage with basing the segmentation on a statistical model is that the uncertainty in the segmentation can be quantified and thus the overall quality of the segmentation can be assessed, e.g. as described by Al-Taie et al.~\cite{Al-Taie2014} However, the segmentation result will only be as good as the model employed and, as pointed out by Van Leemput et al.~\cite{VanLeemput2003}, models providing an improved description of the data is needed. In addition, as for the quantification task, segmentation methods based on models for the intensity like the GMM and PVMM typically misclassify voxels at interfaces in data with more than two phases.

In light of these challenges, we present a novel statistical model for material quantification, describing the joint probability distribution of voxel intensities and gradient magnitudes in 3D tomography data. The proposed approach addresses the above-mentioned problems by incorporating an improved description of interfaces between materials, taking further steps towards a physical model for material quantification and segmentation. Through exploiting gradient magnitude information, it separates interfaces and phase interiors well. Compared to existing models, it allows for additional information about the imaged sample to be extracted and thereby extends the scope for statistical models in materials science. 
Furthermore, the derived distribution model can serve as excellent input to existing probabilistic methods for segmentation. With a better fit to real data, the model proposed in this paper is promising for improving the accuracy of the segmentation and subsequently extracted material parameters. Lastly, our method provides a robust check of segmentation quality, as material parameters measured on the segmentation can be compared to estimated model parameters.

The rest of this paper is structured as follows. In Section \ref{sec:interfaces}, the background on modelling of interfaces in 3D image data is laid out. In Section \ref{sec:method}, we derive the proposed model, and describe the method applied for fitting the model to image data through maximum likelihood estimation (MLE). The implementation (in Python) is publicly available at \url{https://github.com/elobre/bimm}. In Section \ref{sec:results}, we verify and illustrate the diversity of the proposed approach on artificial data, and demonstrate the applicability for experimental data. Finally, we show how the model can be used for segmentation of multi-phased data.

%%%%%%%%%%%%%%%%%%%%%%%%%%%%%%%%%%%%%%%%%%%%%%%%%%%%%%%%%%%%%%%%%%%%%%%%%%%%%%%%%%%%%%%%%%%

\section{Interfaces in 3D Tomography Data}\label{sec:interfaces}

Our model differs from previous mixture models for material quantification in two ways. Firstly, it incorporates Gaussian blurring of interfaces. Secondly, while previous models focused on the intensity distributions only, we also model the distribution of gradient magnitudes. In this section, we show why this is advantageous. 

In 3D tomographic images, edges or interfaces between materials are inevitably blurred due to the finite resolution of the imaging pipeline~\cite{Nuyts2013}. This can be seen as a combination of several effects. Discretization in a digital image leads to what is commonly referred to as the partial volume (PV) effect. A discretization of any interface will lead to mixed-material voxels (mixels) near the interface, with intensities proportional to some weighted average of the attenuation of the adjacent materials. Other steps of the tomographic pipeline will also contribute to blurring of interfaces. The spatial resolution of the imaging system is typically characterized by its point spread function (PSF), which for X-ray CT scanners is commonly modelled as a Gaussian~\cite{Bentzen1983, Nickoloff1985, Kato2013}. The resulting effect in the projection image corresponds to applying a filter with a Gaussian kernel. Afterwards, the 3D tomographic reconstruction and artifact corrections in the image post-processing will also contribute to blurring of the data. In this work, we model the combined effect of the factors limiting the image resolution as Gaussian blurring, resembling the effect of applying a Gaussian filter to sharp interfaces in 3D image data.

\begin{figure}
\includegraphics[width=\mysize\columnwidth]{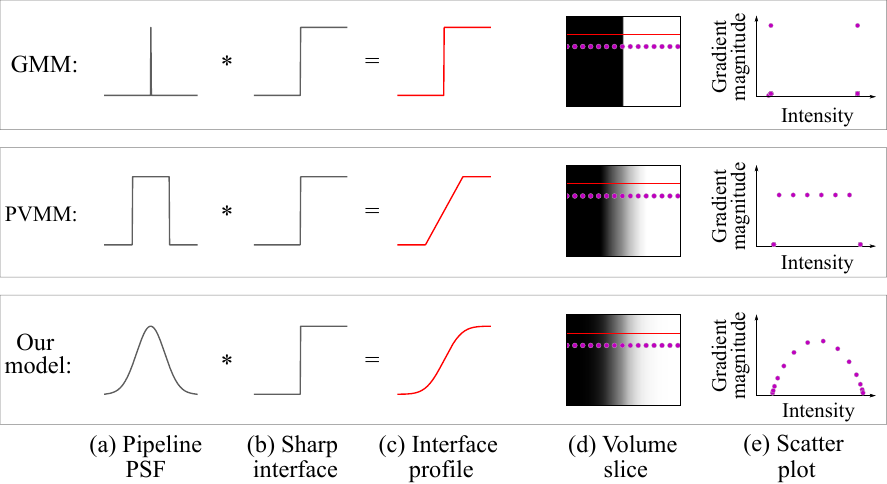}
\caption{  Illustration of the interface models considered in this paper. The interface profiles in column (c) are convolutions between the pipeline PSF approximation (filter kernel) in (a) and a step function (sharp interface) in (b). Column (d) shows a slice through a volume filtered with a kernel corresponding to the PSF in (a). The interface profiles in (c) are the intensity along the red line in the volume slice. The intensity and gradient magnitude of the volume is sampled at the purple dots in (d) and plotted in (e). }\label{fig:interface_models}
\end{figure}

Analytically, this blurring can be described as a convolution between a step function and a Gaussian function with standard deviation $\sigma_b$. This Gaussian can be thought of as the PSF of the full tomographic pipeline. Now, imagine a noise-free data volume containing two materials with intensities $I_i$ and $I_j$. The interface between the two materials is centered at $x=0$ and oriented so that the interface normal points along the $x$-axis. The intensity in a point $(x,y,z)$ is then given by
\begin{linenomath}\begin{equation}\label{eq:I_x}
I(x, y, z) = I_i + (I_j-I_i)  \cdot \frac{1}{2} \left[1+\text{erf} \left(\frac{x}{\sigma_b \sqrt{2}}\right)\right],
\end{equation}\end{linenomath}
where $\text{erf}(u) = \frac{2}{\sqrt{\pi}} \int_0^u \exp (-x' \, ^2) \, \text{d} x'$. In contrast, the PVMM~\cite{Gage1992} can be seen as employing a box-shaped filter kernel, while the GMM assumes a delta function, i.e. infinite resolution. As illustrated in Fig.~\ref{fig:interface_models}, the delta function-approximation to the pipeline PSF results in a sharp interface, the box-approximation gives a linear transition between materials, while the Gaussian approximation gives an interface profile described by the error function (erf). 

\begin{figure}
\includegraphics[width=\mysize\columnwidth]{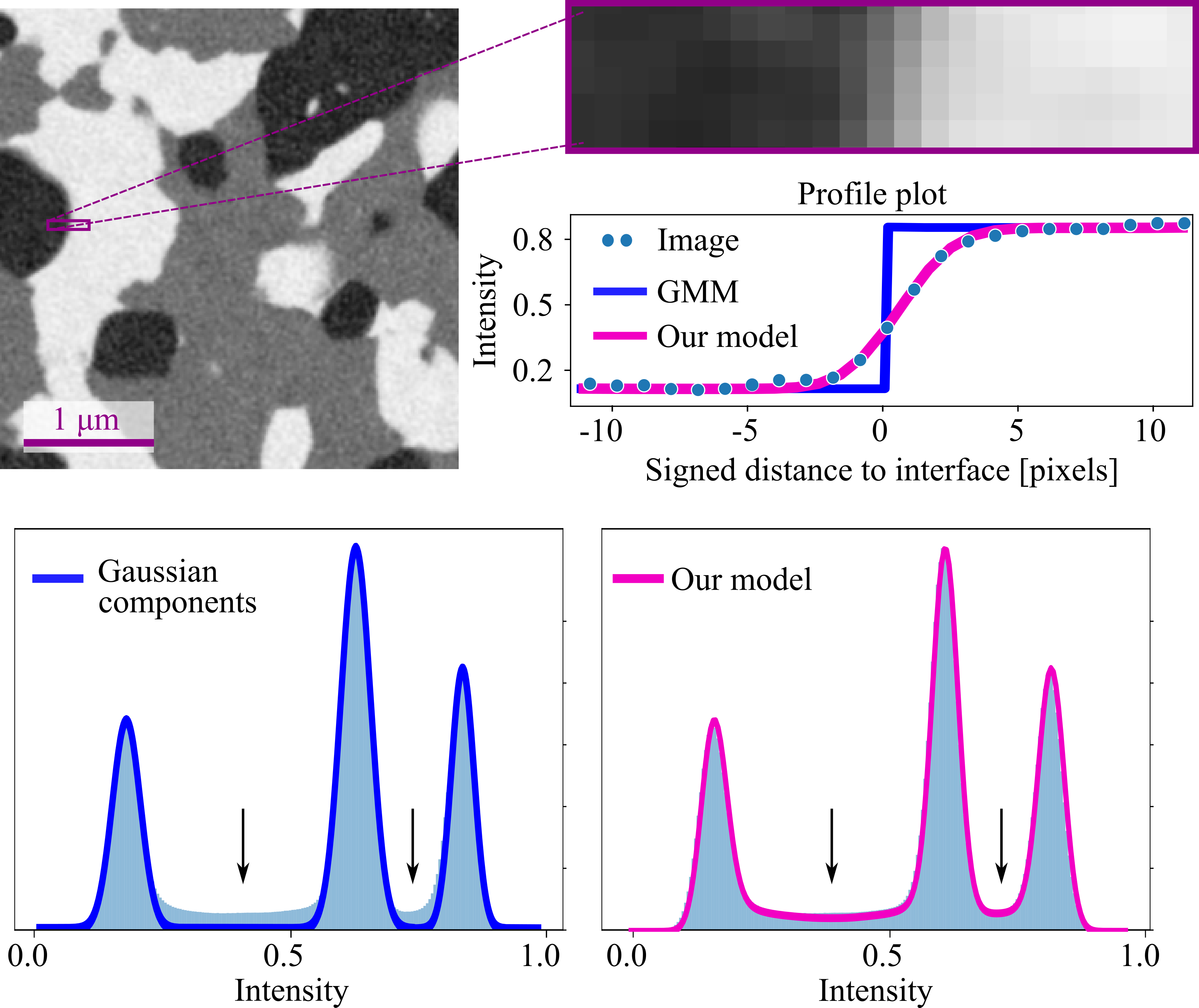}
\caption{ (top) Image data~\cite{DeAngelis2018} and the intensity profile of a black-white interface (dotted line) compared to the GMM (blue step function) and our model (purple error function). (bottom) The two models plotted along with the full dataset histogram. The better description of the interface profile gives a better description of the histogram between the peaks (indicated with arrows).}\label{fig:profile_plot}
\end{figure}

Fig.~\ref{fig:profile_plot} exemplifies how the improved description of the interface profile gives a better model fit to the intensity histogram of experimental image data. While simple Gaussian components model the phase interiors well, showing a good fit for the three histogram peaks, a Gaussian-blurred interface model is needed to model the interface voxels with intensities between the histogram peaks.

\begin{figure*}
\centering
\includegraphics[width=\textwidth]{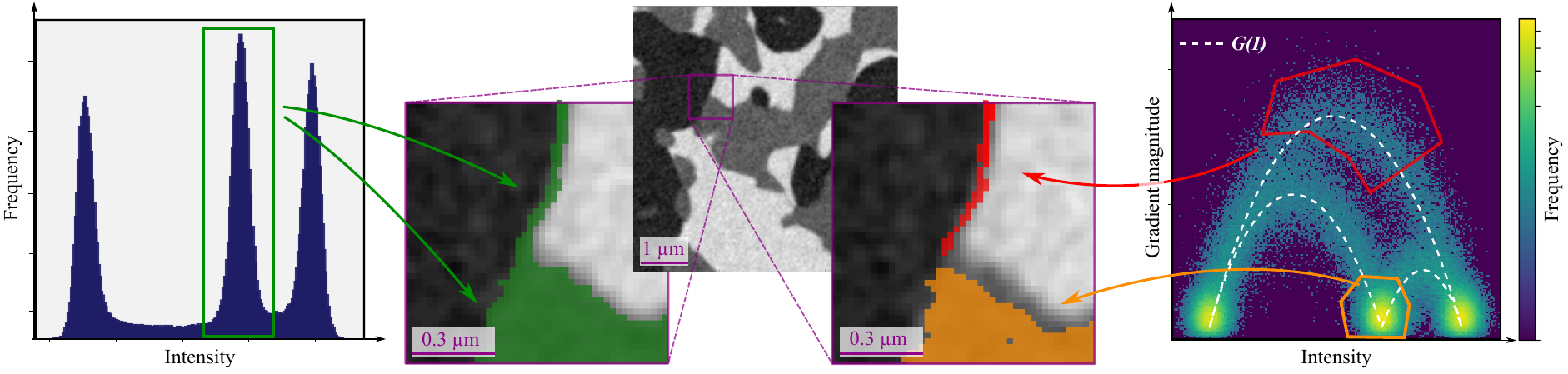}
\caption{Illustration of the advantage of taking gradient information into account. (center) A slice of a 3D dataset containing three different phases and three different types of interfaces~\cite{DeAngelis2018} along with the corresponding intensity histogram (left) and intensity--gradient magnitude histogram (right). In the latter, voxels in the interface between black and white are clearly separated from the grey phase interior, while in the intensity histogram they overlap.  The overlaid stippled white lines indicate $G(I)$ from Eq.~\eqref{eq:G_I}. }\label{fig:BGW_material}
\end{figure*}

In a material with more than two phases, an interface between two phases can have the same intensity as a third phase. As an example, the blurred interface between a black and a white phase will have grey intensities. In a 1D histogram of intensity values, the interface voxels will overlap with the grey phase, as illustrated in Fig.~\ref{fig:BGW_material}. Therefore, accurate classification and quantification of interface regions in multi-phase structures are challenging to perform based on intensity values alone. One approach to addressing this problem is to take gradient information into account. Plotting gradient magnitudes versus intensity values, we see a characteristic arc pattern appearing (Fig.~\ref{fig:BGW_material} (right)). As indicated in the figure, in this 2D intensity--gradient magnitude space, interiors and interfaces with the same intensity values are more easily separated. This fact has previously been exploited in 3D data visualization~\cite{Kindlmann1998} and for material quantification~\cite{Serlie2007}, but no statistical model has been derived. 

The arcs in the 2D intensity--gradient magnitude histogram in Fig.~\ref{fig:BGW_material} (right) are a consequence of Gaussian blurring of sharp interfaces. In contrast, no blurring (the GMM) or a box-filter blurring (the PVMM) would give 2D plots resembling those in Fig.~\ref{fig:interface_models}(e). The white stippled lines overlaid on the 2D histogram in Fig.~\ref{fig:BGW_material} indicate $G(I)$, which is the gradient magnitude expressed in terms of the intensity $I$. We define the gradient magnitude as the Euclidean norm of the gradient of the intensity, 
\begin{linenomath}\begin{equation}%\label{eq:G_I} 
G(x, y, z) = \| \nabla I (x, y, z) \|.
\end{equation}\end{linenomath}
The gradient magnitude is a scalar that is independent of the orientation of the interface. An expression for $G(I)$ can therefore, without a loss of generality, be derived by differentiating $I(x, y, z)$ in Eq.~\eqref{eq:I_x} with respect to $x$ and expressing the result in terms of the intensity $I$,
\begin{linenomath}\begin{equation}\label{eq:G_I} 
G(I) = \frac{\mid I_j-I_i \mid }{\sigma_b \sqrt{2\pi} } \text{exp}\Big(-\Big[ \text{erf}^{-1}\Big(2\frac{I-I_i}{I_j-I_i}-1\Big)\Big]^2 \Big).
\end{equation}\end{linenomath}
From the figure, we see that the arcs agree well with the data, indicating that Gaussian blur is a reasonable model for the interface blurring in this dataset. %Note that because this expression is independent of the choice of coordinate system, it is valid for interfaces at all orientations. 

%%%%%%%%%%%%%%%%%%%%%%%%%%%%%%%%%%%%%%%%%%%%%%%%%%%%%%%%%%%%%%%%%%%%%%%%%%%%%%%%%%%%%%%%%%

\section{Methods}\label{sec:method}

As argued in the previous section, it is advantageous to include gradient information in the model to better distinguish interfaces and phase interiors. The model we seek to derive is therefore the joint probability distribution of voxel intensities $\vu$ and gradient magnitudes $\vv$, denoted $p(\vu,\vv)$. The gradient magnitudes $\vv$ are calculated from neighbouring voxels by central differences. In this section, we first present the main assumptions our model is based upon. Thereafter, we derive the model probability density function (PDF). Lastly, we describe how the model can be fitted to the data using maximum likelihood estimation (MLE). Full details on the model derivation are available in \ref{app:model_derivation}.

\subsection{Model Assumptions}\label{sec:assumptions}

Image artifacts that can affect the distribution of intensities and gradients include beam hardening, striking artifacts from saturated detector pixels, ring artifacts from inconsistent detector sensitivity, illumination drift from beam instability, inaccurate geometric parameters used in the tomographic reconstruction, sample drift during scanning and phase contrast affecting the interface profile. However, with carefully tuned experimental conditions and processing of the data before and after the reconstruction, such artifacts can be greatly reduced~\cite{Boas2012}. For a dataset to be suitable for analysis with the approach proposed in this paper, it is required that image artifacts are corrected for in such a way that the following assumptions hold: 

\begin{itemize}[leftmargin=10pt]\setlength\itemsep{2pt}
    \item \textit{Isotropic, Gaussian image blurring.}\\
    Data acquisition, 3D reconstruction and post\hyp{}processing; each step of the tomographic pipeline will affect the sharpness of the final data. We assume that the combination of effects limiting the image resolution can be modelled as isotropic Gaussian blurring.
    \item \textit{Isotropic, additive and normally distributed noise.} \\
    A certain level of noise will be present stemming from e.g. detector defects or shot noise from the electronics. The noise in the transmitted X-ray photon intensity has a Poisson distribution. However, for large photon counts, a Gaussian distribution is an excellent approximation~\cite{Whiting2002}.
    \item \textit{Correlation between neighbouring voxels.}\\
     In the 3D reconstruction process and data post\hyp{}processing, filters are often applied to reduce the noise. We assume that the filter effects can be sufficiently described as correlation between neighbouring voxels. 
    \item  \textit{Homogeneous materials.}\\
    We assume that each material phase has a constant attenuation at the resolution of our image data, where texture or slight variations in the phase intensity can be modelled as contributing to the Gaussian noise and correlation. For simplicity, we assume the same texture for all phases, and therefore the same correlation for all phases. 
\end{itemize}
Based on these assumptions, we model 3D tomographic image data in terms of the following physical parameters: the standard deviation of a Gaussian-shaped pipeline PSF, denoted $\sigma_b$; the standard deviation of additive Gaussian noise, denoted $\sigma_n$; the correlation between neighbouring voxels, denoted $\rho$; and the mean intensity of the $N$ constituent homogeneous materials, denoted $I_i$, $i=1,2,...,N$.  

\subsection{Modelling Phase Interiors and Interface Regions}\label{sec:interior_interface}

We define the different volume regions in a 3D image dataset as \textit{interface regions} and \textit{phase interiors}. In the former, voxel intensities are significantly affected by the adjacent phase, i.e. the interface blur effect is substantial. In phase interiors, the interface effects are negligible. Our model is a finite mixture model with one component for each of the different volume regions in the dataset at hand,
\begin{linenomath}\begin{equation}\label{eq:p_u_v_split}
     p(\vu,\vv) = \sum_i w_i\, p_i(\vu,\vv) + \sum_i \sum_{j \neq i} w_{ij}\, p_{ij}(\vu,\vv).
\end{equation}\end{linenomath}
Here, the mixture weights $w_i$ and $w_{ij}$ are non-negative and together, they sum to one. $p_i(\vu,\vv) $ denotes the PDF for the interior of material $i$ and $p_{ij}(\vu,\vv) $ denotes the PDF for the interface region between materials $i$ and $j$. As we will see in Section~\ref{sec:params_to_prop}, having separate density components for the different volume regions is key to the quantification procedure. 

More specifically, we define the interface region between materials $i$ and $j$ as all voxels within a distance $d_s\sigma_b$ of the true interface, where $d_s$ is a free parameter in the optimization which typically takes a value between 1 and 3 depending on the size of features in the image data. The region is then centered at the true interface and has a thickness $2 d_s\sigma_b$, reflecting that a larger blur parameter $\sigma_b$ gives a larger region with significant interface effects. With Eq.~\eqref{eq:I_x}, this translates to limits in intensity, $I_{+}$ and $I_{-}$, 
\begin{linenomath}\begin{equation}\label{eq:IminImax}
I_{+, -} = \frac{1}{2}(I_i+I_j) \, \pm \,  \frac{1}{2}|I_j-I_i| \, \text{erf}\left(\frac{d_s}{\sqrt{2}}\right).
\end{equation}\end{linenomath}
Voxels with an expected value within the interval $[I_{-}, I_{+}]$ are modelled as part of the interface region. As an example, with $d_s= 2$, the interface region covers 95 \% of the interval  $[I_i, I_j]$, while $d_s= 2.5$ covers 99 \%. Voxels with an expected value outside the interval $[I_{-}, I_{+}]$ are thus modelled as part of the interior phases. This approximation is better with a large value of $d_s$. At the same time, a small value of $d_s$ will provide a better fit for datasets containing small image features. This is because the interface width $2 d_s \sigma_b$ should be smaller than the diameter of image features or distances between interfaces. An optimal trade-off for a specific dataset is achieved by letting $d_s$ be a free parameter in the optimization.

\subsection{The Blurred Interface Mixture Model}\label{sec:p_u_v}

As argued in Section~\ref{sec:assumptions}, real image data is noisy. The intensity profile $I$ in Eq.~\eqref{eq:I_x} is the \textit{expected} intensity of a voxel, depending on its distance to an interface. We assume that the deviations from $I$ are normally distributed with variance $\sigma_n^2$, i.e. the intensity data $\vu \sim N(I, \sigma_n^2)$. The conditional probability density for $\vu$ given $I$ is then
\begin{linenomath}\begin{equation}\label{eq:p_u_I}
p(\vu \mid I ) = \frac{1}{\sqrt{ 2 \pi \sigma_n^2 }} \text{exp}\left(- \frac{ (\vu - I)^2 } {2\sigma_n^2}\right).
\end{equation}\end{linenomath}
The gradient magnitude $\vv$ is calculated from $\vu$ using central differences. Consequently, $\vv$ has a scaled non-central chi distribution (see \ref{app:dist_v}). The variance of this distribution will depend on the correlation $\rho$ between voxels, capturing the smoothing effect of a noise reduction filter or underlying texture in the materials. The conditional probability density for $\vv$ given $I$ is as follows, 
\begin{linenomath}\begin{equation}\label{eq:p_v_I} 
\begin{split}
p(\vv \mid I) &= \frac{2}{\sigma_n^2(1-\rho)} \sqrt{\frac{ \vv^3}{G(I) }}  \text{exp}\left(-\frac{\vv^2+G(I)^2}{\sigma_n^2(1-\rho)} \right) \\
& I_{\frac{1}{2}}\left(\frac{2 \vv G(I)}{\sigma_n^2(1-\rho)}\right).
\end{split}
\end{equation}\end{linenomath}
Here, $I_{\frac{1}{2}}$ is the modified Bessel function of first kind. Note that $G(I)$ appears in this expression. This is because the non-centrality parameter of the distribution can be written in terms of the central differences approximation of $G(I)$. Details of the derivation can be found in \ref{app:dist_v}. The joint probability distribution for $\vu$ and $\vv$, $p(\vu,\vv)$, can now be found from Eq.~\eqref{eq:p_u_I} and Eq.~\eqref{eq:p_v_I} through marginalization,
\begin{linenomath}\begin{equation}\label{eq:p_u_v_integral}
 p(\vu,\vv) =\int p(\vu \mid I)\, p(\vv \mid I)\, p(I) \,\, \text{d}I .
\end{equation}\end{linenomath}
Here, $p(I)$ is the \textit{density of expected intensities} $I$, reflecting the amount of voxels in the volume with expected value $I$. $p(I)$ depends on the density of voxels at a distance $x$ from the interface, $p(x)$, which is in turn determined by the geometry (volume fractions, interface area and curvature) of the imaged object. While Eq.~\eqref{eq:p_u_I} and Eq.~\eqref{eq:p_v_I} are valid for the whole data volume, $p(I)$ is split in phase interior and interface regions as defined in the previous section,
\begin{linenomath}\begin{equation}\label{eq:p_I}
 p(I) =  \sum_i w_i\,  p_i(I) + \sum_i \sum_{j \neq i} w_{ij}\,  p_{ij}(I).
\end{equation}\end{linenomath}
Assuming a uniform distribution of voxels along the interface normal in the interface region, $x \sim U([-d_s\sigma_b, d_s\sigma_b])$, the density can be found through a change of variables,
\begin{linenomath}\begin{equation}\label{eq:p_ij_I}
 p_{ij}(I) = p(x)\, \left\vert \frac{\partial I(x)}{\partial x} \right\vert^{-1} = \frac{1}{2d_s\sigma_b}\frac{1}{G(I)}.
\end{equation}\end{linenomath}
In phase interiors, $G(I) \rightarrow 0$ so $p(I)$ can be approximated by a delta function,  
\begin{linenomath}\begin{equation}\label{eq:p_i_I}
    p_i(I) = \delta(I-I_i).
\end{equation}\end{linenomath}
Combining Eq.~\eqref{eq:p_I} and Eq.~\eqref{eq:p_u_v_integral} we can find the model components in Eq.~\eqref{eq:p_u_v_split}. For the phase interiors, we get 
\begin{linenomath}\begin{equation}\label{eq:p_i_u_v}
\begin{split}
p_i(\vu,\vv) &= \int p(\vu \mid I)\, p(\vv \mid I)\, p_i(I) \,\, \text{d}I \\
&=  \frac{1}{\sqrt{ 2 \pi \sigma_n^2 }} \text{exp}\left( -  \frac{(\vu - I_i )^2} {2\sigma_n^2} \right)  \frac{2 \vv^2}{\Gamma(3/2) \sigma_n^3(1-\rho)^{3/2}}\\
&\text{exp}\left(-\frac{\vv^2}{\sigma_n^2(1-\rho)}\right),
\end{split}
\end{equation}\end{linenomath}
where $\Gamma$ is the gamma function. The integral defining the interface component is, however, more complicated: 
\begin{linenomath}\begin{equation}\label{eq:p_ij_u_v}
\begin{split}
p_{ij}(\vu,\vv) &= \int p(\vu \mid I)\, p(\vv \mid I)\, p_{ij}(I) \,\, \text{d}I \\
&= \int_{I_{-}}^{I_{+}} \frac{1}{\sqrt{ 2 \pi \sigma_n^2 }} \text{exp}\left( -   \frac{ (\vu - I )^2} {2\sigma_n^2}  \right)  \frac{2}{\sigma_n^2(1-\rho)} \\
&\sqrt{\frac{ \vv^3}{G(I) }}  \text{exp}\left(-\frac{\vv^2+G^2(I)}{\sigma_n^2(1-\rho)} \right) \, I_{\frac{1}{2}}\left(\frac{2 \vv G(I)}{\sigma_n^2(1-\rho)}\right)\\
 &\quad \frac{ \sqrt{2\pi}}{2 d_s ( I_j-I_i) } \text{exp}\left(\left[ \text{erf}^{-1}\left(2\frac{I-I_i}{I_j-I_i}-1\right)\right]^2 \right) \, \text{d}I,
\end{split}
\end{equation}\end{linenomath}
where $I_{-}$ and $I_{+}$ are defined by Eq.~\eqref{eq:IminImax}. This integral is not known in closed form, but its value can be approximated using the Monte Carlo method~\cite{Metropolis1949} (details in \ref{app:MCgrad}). The full model is thus expressed by Eq.~\eqref{eq:p_u_v_split} with $p_i(\vu, \vv)$ and $p_{ij}(\vu, \vv)$ given by Eq.~\eqref{eq:p_i_u_v} and Eq.~\eqref{eq:p_ij_u_v}.

\subsection{The Model for Intensity Data Only}\label{sec:p_u}

We refer to the model $p(\vu,\vv)$ derived above as a 2D model, as it is a combined model for intensities and gradient magnitudes in 3D image data. However, a 1D equivalent can be derived, modelling the intensity data only; namely the marginal distribution $p(\vu)$. Such 1D version of the model is useful as it enables a more direct comparison to the 1D models GMM and PVMM. The $\vv$ dependency is removed from Eq.~\eqref{eq:p_u_v_integral},
\begin{linenomath}\begin{equation}\label{eq:p_u}
 p(\vu) =\int p(\vu \mid I)\, p(I) \,\, \text{d}I = \sum_i w_i\, p_i(\vu) + \sum_i \sum_{j \neq i} w_{ij}\, p_{ij}(\vu).
\end{equation}\end{linenomath}
The 1-D interior and interface model components are then as follows,
\begin{linenomath}\begin{equation}\label{eq:p_i_u}
p_i(\vu) = \frac{1}{\sqrt{ 2 \pi \sigma_n^2 }} \text{exp}\left( -  \frac{(\vu - I_i )^2} {2\sigma_n^2} \right) ,
\end{equation}\end{linenomath}
\begin{linenomath}\begin{equation}\label{eq:p_ij_u}
\begin{split}
p_{ij}(\vu) = \int_{I_{-}}^{I_{+}} \frac{1}{\sqrt{ 2 \pi \sigma_n^2 }} \text{exp}\left( -  \frac{( \vu - I)^2 } {2\sigma_n^2}  \right) \, \frac{ \sqrt{2\pi}}{2 d_s ( I_j-I_i) }\\
\text{exp}\left(\left[ \text{erf}^{-1}\left(2\frac{I-I_i}{I_j-I_i}-1\right)\right]^2 \right) \, \text{d}I.
\end{split}
\end{equation}\end{linenomath}
Note that while the shape of the Gaussian-blurred interface is part of this equation through the last exponential term, this 1-D model is independent of the interface blur parameter $\sigma_b$. It is therefore not suitable for assessing the data resolution, calculating interface areas or for segmentation of multi-phased material data. For this, we need the 2D model $p(\vu,\vv)$ (Eq.~\eqref{eq:p_u_v_split}), which depends on $\sigma_b$ through $G(I)$.

In comparison, the GMM consists of the interior components in Eq.~\eqref{eq:p_i_u} only, and no interface components. The PVMM assumes an equal probability of all materials in the mixed-intensity interface voxels, i.e. a uniform intensity distribution $p_{ij}(I) = \frac{1}{I_j - I_i}$ at the interfaces, 
\begin{linenomath}\begin{equation*}\label{eq:p_ij_u_pvgmm}
p_{ij}^{\mbox{\tiny PVMM}}(\vu) = \int_{I_{i}}^{I_{j}} \frac{1}{\sqrt{ 2 \pi \sigma_n^2 }} \text{exp}\left( - \frac{ (\vu - I )^2} {2\sigma_n^2} \right)  \, \frac{ 1}{ I_j-I_i } \text{d}I.
\end{equation*}\end{linenomath}
This integral for the PVMM interface component does have an analytic solution but for the sake of numerical stability in the implementation, MC integration is used in the comparison below.

\subsection{From Model Parameters to Physical Properties}\label{sec:params_to_prop}

The model can be exploited for measurements of material structure properties. When fitted to a dataset, the resulting model parameters provide estimates of physical properties of the imaged sample. For example, the material density of phase $i$ is directly proportional to the model parameter $I_i$, and the noise level is equal to the model parameter $\sigma_n$. Estimates of volume fractions, interface areas and resolution can be found as follows. The interface component weight $w_{ij}$ is proportional to the volume of the interface region between phases $i$ and $j$, with the thickness of the region defined as $2 d_s \sigma_b$. The interface area $A_{ij}$ can therefore be calculated as
\begin{linenomath}\begin{equation}\label{eq:interface_area}
A_{ij} = V \frac{w_{ij}}{2 d_s \sigma_b}.
\end{equation}\end{linenomath}
Here, $V$ is the dataset volume. The interface component is symmetrical, so the volume fractions $V_i$ are given by

\begin{linenomath}\begin{equation}\label{eq:volume_fractions}
V_i = w_i + \frac{1}{2} \sum_{j \neq i} w_{ij}.
\end{equation}\end{linenomath}

\begin{figure}
\includegraphics[width=\mysize\columnwidth]{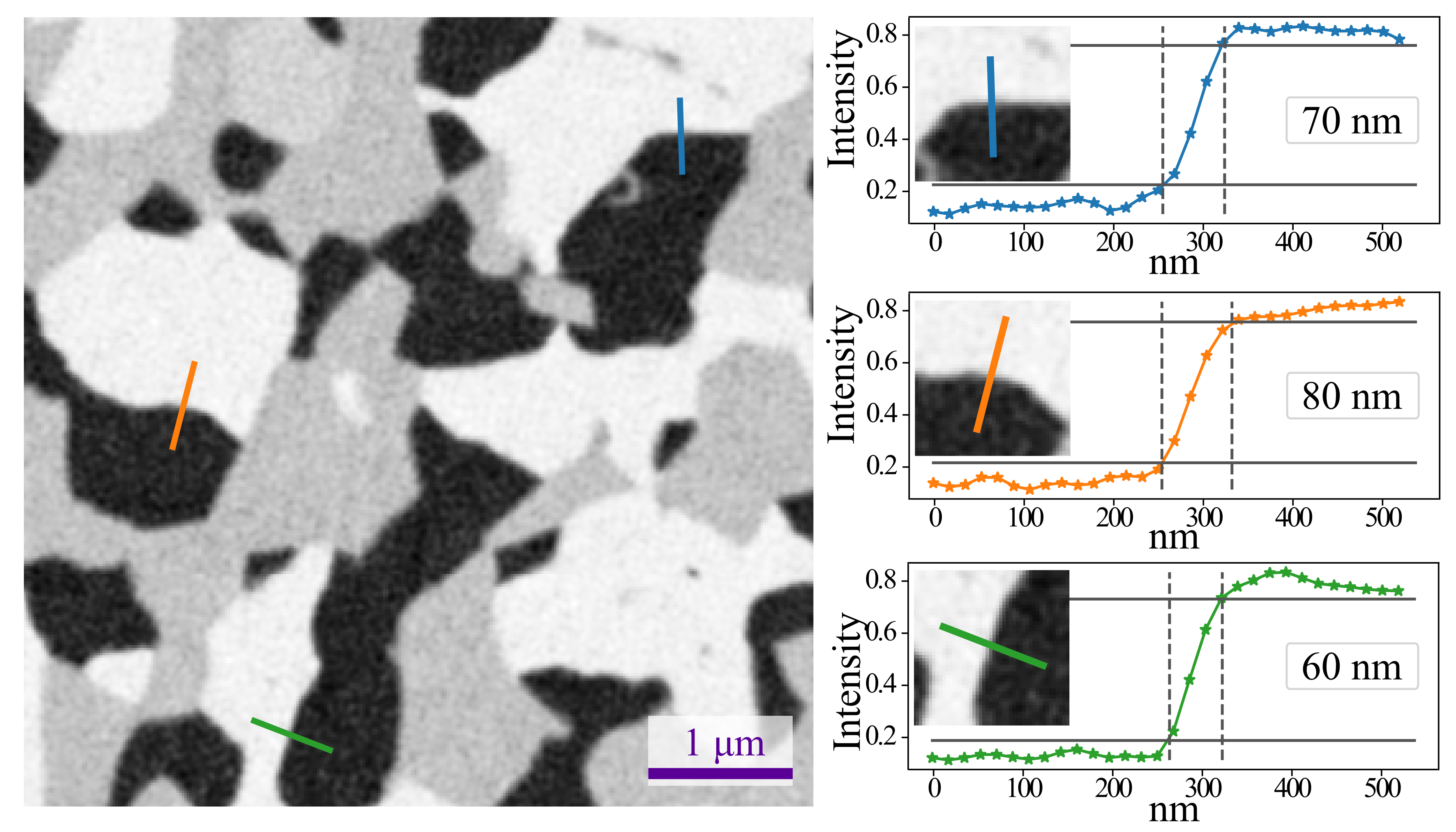}
\caption{According to the 10 \%--90 \%  criterion, the resolution of the data is the width where an intensity profile goes from 10 \% to 90 \%, indicated here with dashed lines. The result using manually extracted profiles can however vary greatly. Data from \cite{DeAngelis2018}. }\label{fig:pristine_profiles}
\end{figure}

A common way to estimate the resolution in image data is to look at the \textit{edge response}, typically described in terms of the 10 \%--90 \%  criterion~\cite{Holler2014}. This is defined as the width where the intensity profile across an interface goes from 10 \% to 90 \%, as illustrated in Fig.~\ref{fig:pristine_profiles}. The dashed vertical lines indicate where the interface profile intersects the 10 \% and 90 \% intensity levels (horizontal solid lines). The resolution according to the 10 \%--90 \% criterion is the distance between these dashed lines which we denote $res_{\text{10--90}}$. Inverting Eq.~\eqref{eq:I_x} to find $x(I)$ allows us to estimate the resolution directly from the model parameter $\sigma_b$,
\begin{linenomath}\begin{equation*}%\label{eq:x_I} 
x(I) = \sigma_b\sqrt{2} \, \text{erf}^{-1}\left(2\frac{I-I_i}{I_j-I_i}-1\right),
\end{equation*}\end{linenomath}
\begin{linenomath}\begin{equation}\label{eq:1090}
res_{\text{10--90}} = | x(I_{90\%} ) - x(I_{10\%}) | =   2 \sigma_b \sqrt{2} \, \text{erf}^{-1}(0.8).
\end{equation}\end{linenomath}

\subsection{Estimating Model Parameters }\label{sec:model_fit}

The derived model, $p(\vu,\vv)$ in Eq.~\eqref{eq:p_u_v_split}, can be fitted to the data using maximum likelihood estimation (MLE). This method finds the model parameters $\vtheta_{max}$ which minimize the \textit{negative log-likelihood},
\begin{linenomath}\begin{equation}\label{eq:L}
 L(\vtheta)=  -\log \prod_{m=1}^M  p(u_m,v_m)= -\sum_{m=1}^M \log p(u_m,v_m).
\end{equation}\end{linenomath}
We use the Adadelta optimizer~\cite{Zeiler2012}, a variant of gradient decent with adaptive learning rates. We use \textit{mini-batches}, meaning that only a few randomly sampled datapoints are used to compute the gradient estimate in each iteration, saving memory and processing time. The size of these mini-batches is one important optimization hyperparameter, referred to below as the batch size. The procedure is implemented in Python, making use of automatic differentiation via the package Pytorch. More details are provided in \ref{app:gradient_descent}  and the code.\footnote{\label{foot:code} \url{https://github.com/elobre/bimm}}

Care must be taken when computing $\nabla_{\vtheta} L$ because of the Monte Carlo (MC) integration used to approximate the interface components of $p(\vu,\vv)$ (Eq.~\eqref{eq:p_ij_u_v}). The \textit{reparametrization trick}~\cite{Kingma2014} is needed to ensure correct gradient estimates. For more details, see \ref{app:MCgrad} and the code.\textsuperscript{\ref{foot:code}} The number of MC samples used in the integration is another important hyperparameter affecting the speed and accuracy of the model fitting, as discussed in Section~\ref{sec:discussion_parameters}.

%%%%%%%%%%%%%%%%%%%%%%%%%%%%%%%%%%%%%%%%%%%%%%%%%%%%%%%%%%%%%%%%%%%%%%%%%%%%
\section{Results and Discussion}\label{sec:results}

In the following, we will refer to the model derived above as the \textit{blurred interface mixture model} (BIMM). As explained, the 2D model $p(\vu, \vv)$ (Eq.~\eqref{eq:p_u_v_split}) describes the combined intensity and gradient magnitude distribution, while the 1D version $p(\vu)$ (Eq.~\eqref{eq:p_u}) models the intensity distribution only. We will refer to these as the BIMM-2D and the BIMM-1D, respectively.

In this section, we first demonstrate on artificial data how our model can be used to extract material parameters, and investigate the difference in performance compared to the GMM and the PVMM. We illustrate the advantages of including gradient information in the model. Next, we look at experimentally obtained data, using the BIMM-2D to extract volume fractions, resolution and interface areas in data with three phases. Finally, we show how the BIMM-2D can be exploited for segmentation.

%optimization parameters
The results presented below are obtained as explained in Section~\ref{sec:model_fit}, using a batch size of 50 and 1000 MC samples. 2000 iterations of the optimization algorithm were run, and the final fitted model parameters are calculated as the mean of the last 500 iterations. See Section~\ref{sec:discussion_parameters} for further discussions on the choice of optimization parameters. 

%%CI runs
The reproducibility of the model parameter estimation procedure is analyzed by fitting the model to 30 independent random sub-samples of the data. This allows us to calculate confidence intervals for the material parameters estimated by the model with an approach resembling bootstrapping. Note that these confidence intervals are only reflecting the variance of estimated material parameters for repeated model fittings, and not the accuracy of the estimate.

\subsection{Estimating Material Parameters in Artificial Data}\label{sec:results_art_data}

\begin{figure}
     \centering
      \captionsetup[subfigure]{justification=centering}
     \subfloat[ \\ $V_1=0.1$, $V_2=0.9$]{\includegraphics[width=\mysizethird\columnwidth]{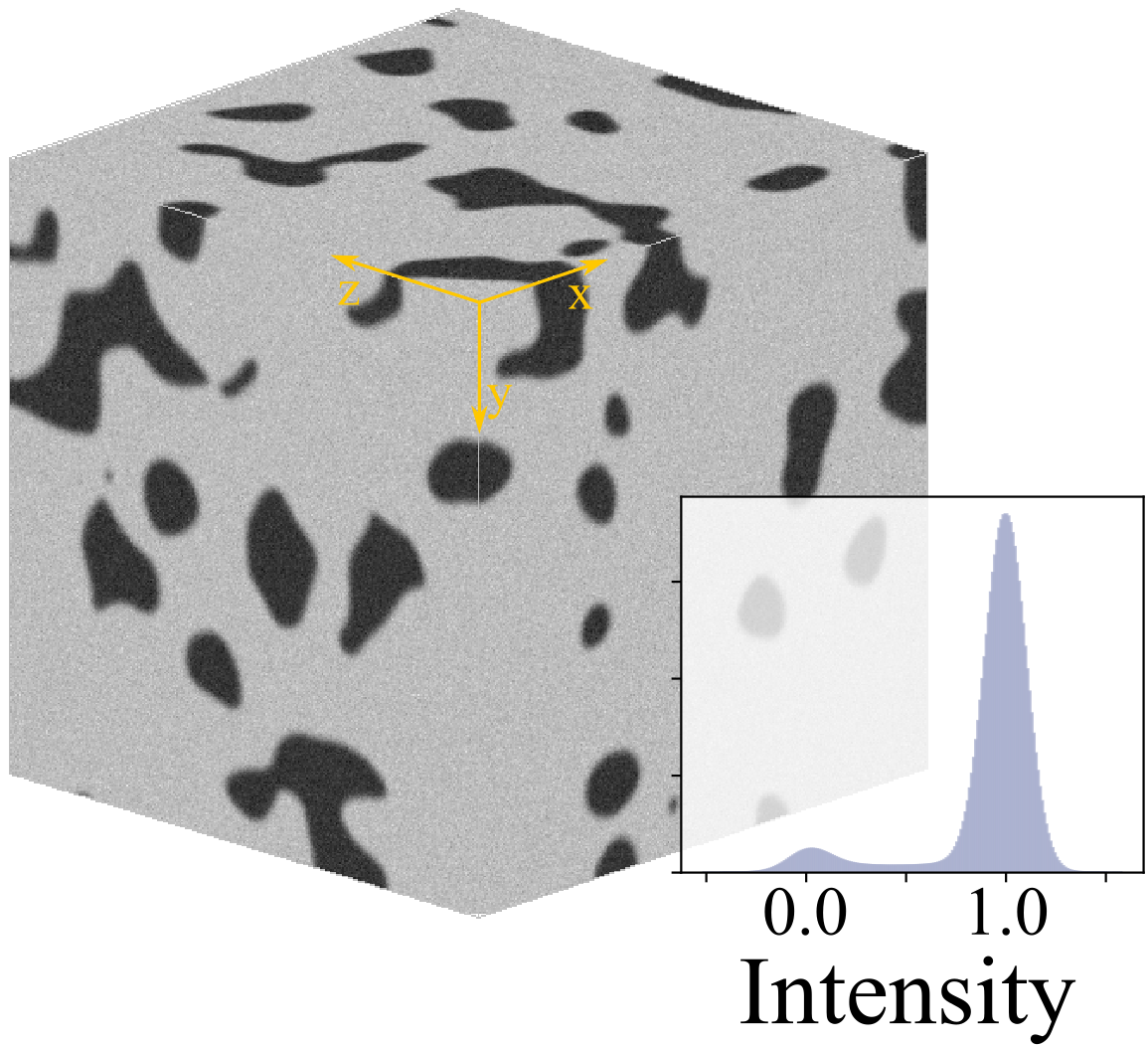}\label{fig:art_data_0109}}
     ~
     \subfloat[ \\ $V_1=0.5$, $V_2=0.5$]{\includegraphics[width=\mysizethird\columnwidth]{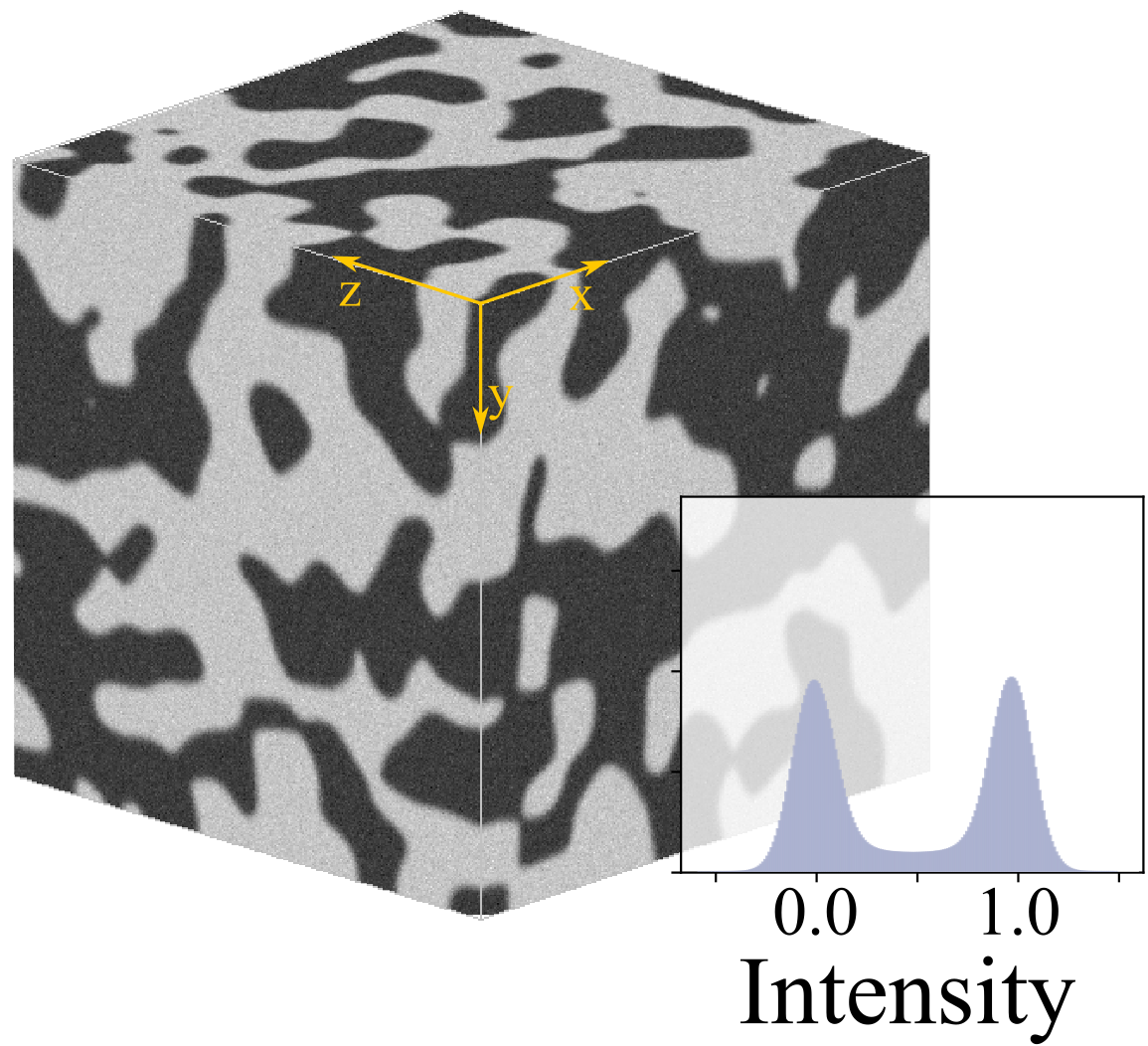}\label{fig:art_data_0505}}
     ~
     \subfloat[ \\ $V_1=0.9$, $V_2=0.1$]{\includegraphics[width=\mysizethird\columnwidth]{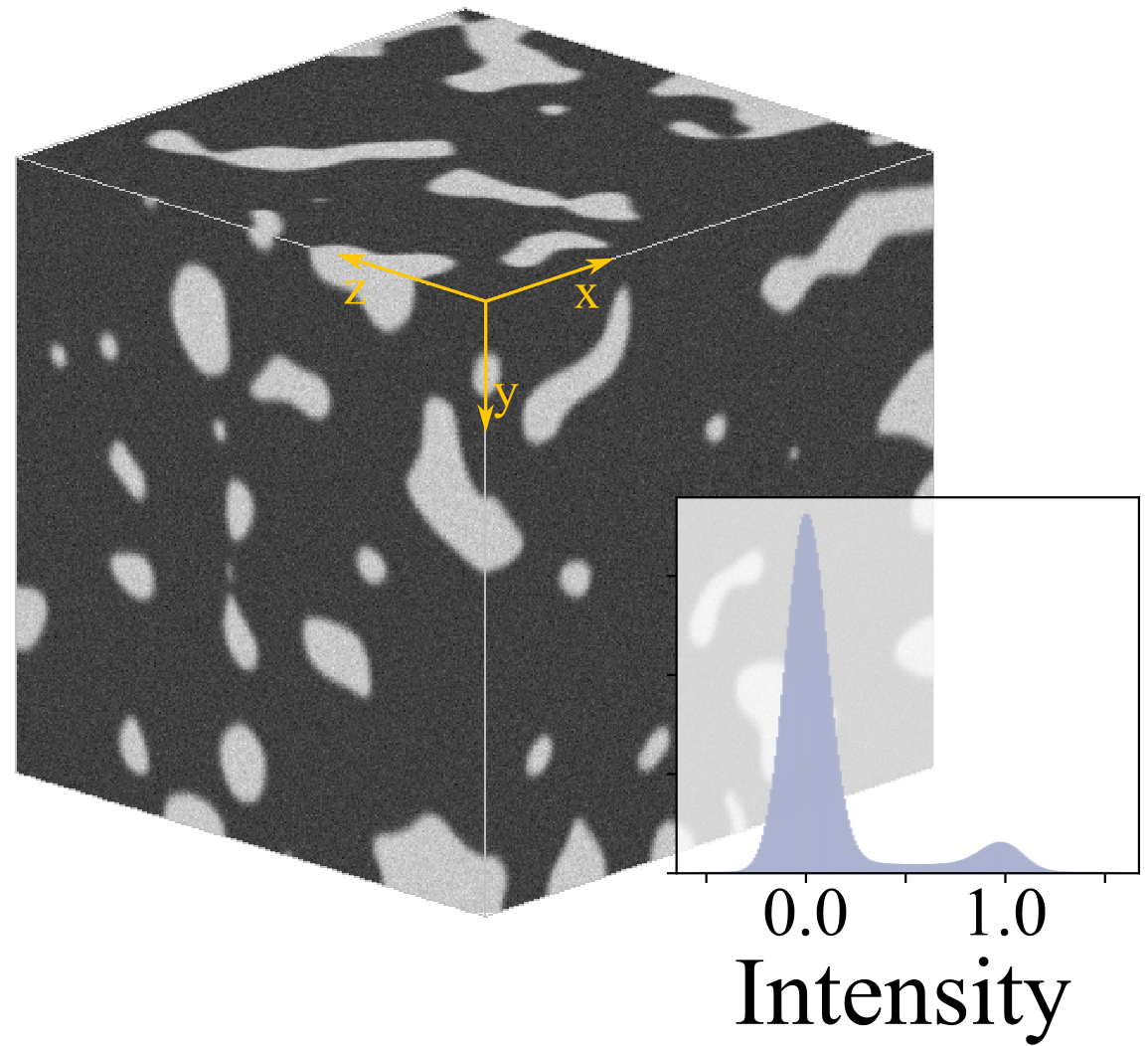}\label{fig:art_data_0901}}
     \caption{Three of the artificial datasets along with corresponding intensity histogram. The full artificial data series contains 19 volumes with the volume fraction $V_1$ ranging between 0.05 and 0.95. }
     \label{fig:art_data}
\end{figure}

Artificial datasets were generated, representing a material with two phases (pores and solid material). In total, 19 datasets with dimensions 500 x 500 x 500 voxels were generated with the pore volume fraction $V_1$ ranging between 0.05 and 0.95. Fig.~\ref{fig:art_data} shows three of the volumes along with corresponding intensity histograms. The datasets were generated as illustrated in the Supplementary Material, Fig.~S1. First, standard normally distributed noise was blurred with a Gaussian filter (standard deviation 20) and thresholded at different intensity levels to give a binary volume with the desired volume fractions and a ``blobby" texture. The ``ground truth" interface area and volume fractions were measured in this volume. Thereafter, a Gaussian filter with standard deviation $\sigma_{b} = 3$ voxels was applied. The filter size was 25 x 25 x 25 voxels, truncated at 4 standard deviations. The binary volume before blurring contained values 0 and 1 representing the two material phases, so $I_{1}=0$ and $I_{2}=1$. Normally distributed (Gaussian) noise was then added with zero mean and standard deviation $\sigma_{n} = 0.1$, i.e. 10 \% noise. No subsequent filter was applied so the correlation parameter $\rho = 0$. %\ref{fig:art_data_gen}

\subsubsection{Estimating Volume Fractions and Phase Intensities}

\begin{figure}
\includegraphics[width=\mysize\columnwidth]{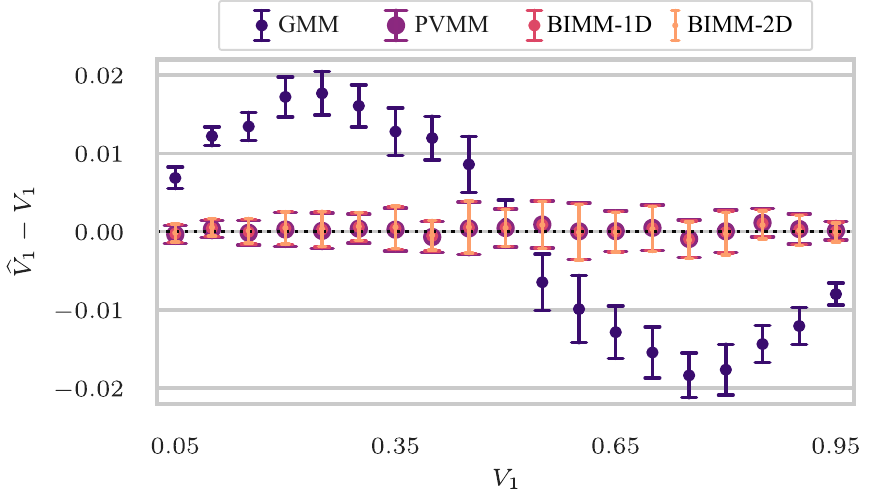}
\caption{Deviation from ground truth value $V_1$ in estimated volume fraction $\widehat{V}_1$ for all artificial datasets. The error bars indicate $\pm$ one standard deviation. } \label{fig:Vs}
\end{figure}

%%%% VOLUME FRACTIONS
To quantify the amount of each material phase is a common task when characterizing a sample from 3D image data. Fig.~\ref{fig:Vs} shows how the BIMM (1D and 2D) compares to the PVMM and GMM in estimating volume fractions in the artificial data. The volume fractions are calculated from the model weights $w$ using Eq.~\eqref{eq:volume_fractions}. The difference between the model estimate $\widehat{V}_1$ and the ground truth value $V_1$ is plotted for each of the datasets. Since the volume fractions sum up to one, the deviation for the other model component, $\widehat{V}_2-V_2$, is identical but with opposite sign. The results for the PVMM and the BIMM (1D and 2D) are comparable, and similar for all datasets, with an average absolute deviation of 0.001. The GMM gives less accurate estimates, with a maximum deviation of 0.018. 

The graph showing the GMM results has an interesting S-shape. In the extreme cases, $V_1=0$ and $V_1=1$, there is only one phase and no interface present in the data so the GMM should be exact. For the dataset with $V_1=V_2=0.5$, the interface voxels are split equally between the two model components, also giving an accurate result. This is however not due to a good model fit but rather due to the fit being equally poor for both peaks in the intensity distribution. For the datasets in between, we see that the GMM is not able to model the interface voxels properly, resulting in worse volume fraction estimates and the S-shaped deviation graph.

\begin{figure}
\includegraphics[width=\mysize\columnwidth]{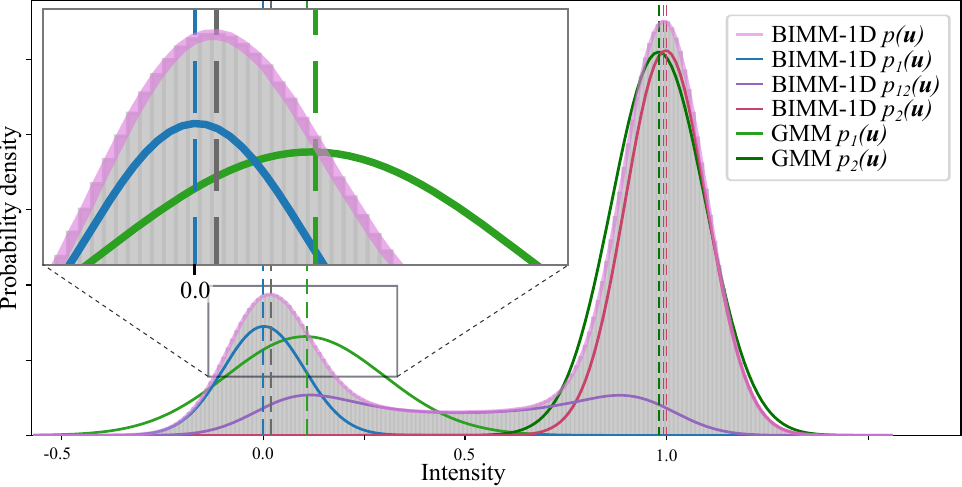}
\caption{Using peaks to estimate phase intensities ($I_1=0$ and $I_2=1$ for this data). Dashed vertical lines indicate the peak of the function in the same color. The dashed grey lines indicate the position of the histogram peaks. Inset: The peak of the BIMM-1D interior component (blue) gives a better estimate of $I_1$ compared to the GMM (green) and the histogram peak (grey). The total BIMM-1D (semi-transparent pink: sum of the three model components) fits the data well. } \label{fig:dicussion_peaks0307}
\end{figure}

Histogram peak positions are often used as an estimate of the mean intensity of the phases, used e.g. for finding thresholds for segmentation or estimating the material density. Fig.~\ref{fig:dicussion_peaks0307} shows the normalized intensity histogram of one artificial dataset with ground truth phase intensities $I_1=0$ and $I_2=1$ along with the fitted BIMM-1D and GMM components. The dashed lines help us to compare the positions of the histogram peaks and the model component peaks. We see that the peak of the BIMM-1D interior components are closer to the ground truth values compared to the peaks of the histogram and the GMM model components, which are biased upwards for $I_1$ and downwards for $I_2$ because of the interface voxels with intensities between $I_1$ and $I_2$. We also see that the bias is larger for the smaller model component. This illustrates that histogram and GMM peak positions in general are not indicative of the true phase intensity value, as the peak positions are shifted by the contribution of the voxels near the interfaces. By splitting the BIMM into phase interior components and interface components, a more accurate phase intensity estimate is achieved.

\begin{figure}
\includegraphics[width=\mysizehalf\columnwidth]{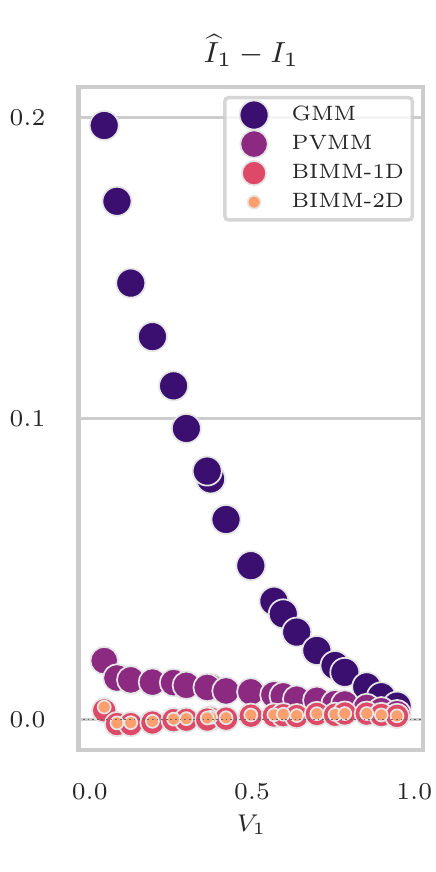}
\includegraphics[width=\mysizehalf\columnwidth]{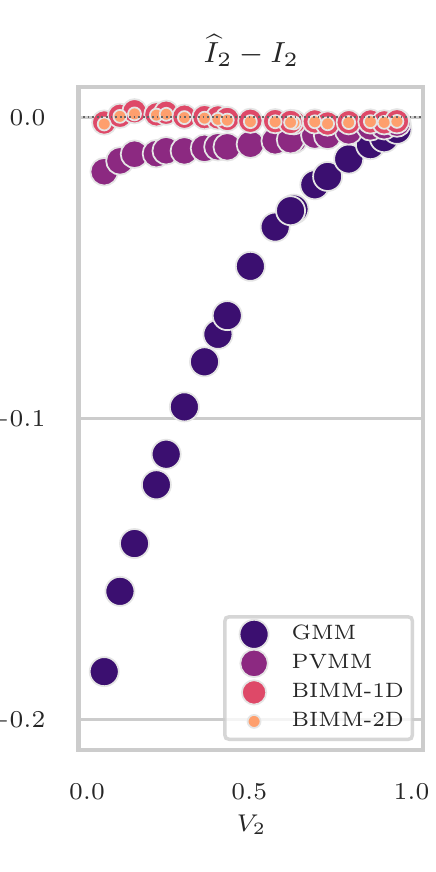}
\caption{Deviation from ground truth values for $\widehat{I}_1$ and $\widehat{I}_2$ plotted against volume fractions $V_{1}$ and $V_{2}$ for all artificial datasets. The results shown are the mean of 30 model fittings. Error bars are left out for clarity - the standard deviations are, on average, \num{5e-3} for the GMM and \num{3e-3} for the other models.  } \label{fig:discussion_peaks_I1I2dev_}
\end{figure}

Fig.~\ref{fig:discussion_peaks_I1I2dev_} shows how the BIMM (1D and 2D) compares to the GMM and the PVMM in estimating the phase intensities for the artificial datasets. The plot shows the difference between model estimate and ground truth values plotted against the volume fraction of that component, $\widehat{I}_1 - I_1$ vs. $V_1$ and $\widehat{I}_2-I_2$ vs. $V_2$. For the GMM and the PVMM, a systematic trend is seen where the deviation from ground truth is higher the smaller the volume fraction is. We see a consistent overestimation of the low intensity phase and an underestimation of the high intensity phase. Overall, the results for BIMM-1D and BIMM-2D are very similar and show a better performance than both the GMM and the PVMM, with absolute deviations below 0.2 \% for all volume fractions. 

It is not surprising that the GMM performs poorly as it does not attempt to model the interfaces in the data. For the phase intensity estimation, the error is small for one phase if its volume fraction is very large, but in return, the error is then large for the other phase with a small volume fraction. For the volume fraction estimation, the GMM errors are typically small for data with similar phase fractions, and would also be small for datasets with a small fraction of interface voxels. However, as shown, the errors in estimated phase intensities and volume fractions are always systematically biased and consequently any quantitative measurements based on a GMM fit will contain this bias.

\subsubsection{Estimating Resolution and Interface Areas}

In the previous section, we saw that the BIMM-1D and the BIMM-2D gave very similar results when estimating volume fractions and phase intensities. The advantage of the BIMM-2D is that we can obtain additional information through the model parameter $\sigma_b$ which reflects the level of interface blurring.

First, we look at how the estimated BIMM-2D parameters compare to the intensities, noise level and blur level used when generating the artificial data. The values are listed in Table~\ref{tab:data_gen_vs_model_est}, averaged over all datasets. See the Supplementary Material, Fig.~S2 for a plot of the results for individual datasets, showing only a minor dependence on the volume fraction ratio. All model estimates are seen to agree well with the ground truth parameters used to generate the data. %\ref{fig:params_art}

\begin{table} %% batch size 50
\caption{Parameters used when generating the artificial data (ground truth) vs. parameters estimated using the BIMM-2D (mean and standard deviation of results for all datasets).}\label{tab:data_gen_vs_model_est}
%\begin{ruledtabular}
\centering
\begin{tabular}{llll}
\toprule
  \multicolumn{2}{l}{ Data generation} & \multicolumn{2}{l}{Model estimate} \\
 \hline\\
 $\sigma_b$ & 3 &  $\widehat{\sigma}_b$ & $3.04 \pm 0.04 $ \\
 $\sigma_n$ & 0.1 & $\widehat{\sigma}_n$ & $0.1009 \pm 0.0005 $\\
$\rho$ & 0 & $\widehat{\rho}$ &  $0.016 \pm 0.009 $\\
 $I_1$ & 0 & $\widehat{I}_1$ & $0.001 \pm 0.002 $\\
$I_2$ & 1 & $\widehat{I}_2$ & $0.999 \pm 0.002$ \\
\bottomrule \\
\end{tabular}
%\end{ruledtabular}
\end{table}

\begin{figure}
\begin{minipage}[t]{0.47\columnwidth}
  \includegraphics[width=\textwidth]{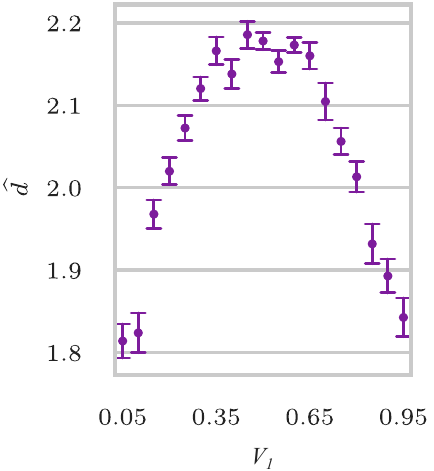}
  \caption{$\widehat{d}$ estimated by the BIMM-2D for all artificial datasets. This parameter determines the width of interface region, $2 d_s \sigma_b$. The average standard deviation is 0.03.}
  \label{fig:d_art}
\end{minipage}%
\hfill
\begin{minipage}[t]{0.47\columnwidth}
  \includegraphics[width=\textwidth]{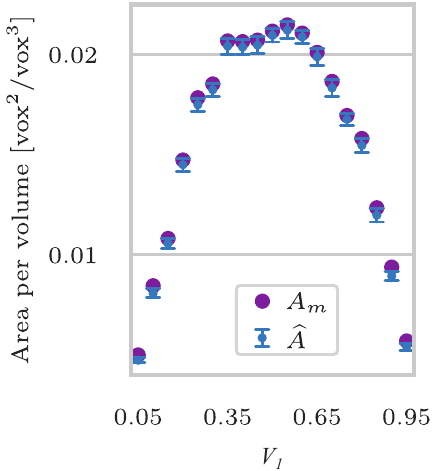}
  \caption{Interface area for all artificial datasets. $\widehat{A}$ are results using the BIMM-2D while $A_{m}$ are calculated using marching cubes.}
  \label{fig:res_area}
\end{minipage}
\end{figure}

Due to the way the artificial datasets are generated, the size of image features (pore sizes) varies with the volume fractions. As seen in Fig.~\ref{fig:art_data}, the volume with $V_1=0.5$ has larger image features compared to the volumes with $V_1=0.1$ (small black pores) and $V_1=0.9$ (small grey particles). As explained in Section~\ref{sec:interior_interface}, the image feature sizes influence what is the optimal width of the interface region, defined as $2d_s\sigma_b$. In volumes with small pores or material inclusions, a narrow region gives a better model fit. This is reflected in the results for $\widehat{d}$, plotted in Fig.~\ref{fig:d_art} for all artificial datasets. The plot shows that $\widehat{d}$ is larger (around 2.15) for volumes with large image features ($V_1 =$ \SIrange{0.3}{0.7}{}), and smaller (down to 1.8) for volumes with smaller image features (smaller and larger value of $V_1$). In contrast, other model parameters do not show a similar dependence on the image feature size (see Supplementary Material, Fig.
~S2).%~\ref{fig:params_art} 

%%% RES 10-90
As explained in Section~\ref{sec:params_to_prop}, the resolution of 3D image data is often stated in terms of the 10 \%--90 \% criterion. This can be found by measuring the intensity profile along the normal of an interface, preferably in many locations to get a good statistics. If done manually, this is a challenging and time consuming task. Manually measuring the 10 \%--90 \% criterion at 5 locations in one of the artificial datasets gave a resolution of 8 voxels. The resolution calculated using $\widehat{\sigma}_b$ from Table~\ref{tab:data_gen_vs_model_est} and Eq.~\eqref{eq:1090} agrees well with this, $res_{\text{10--90}} = 7.8 \pm 0.1$ voxels. 

%%% INTERFACE AREA 
Fig.~\ref{fig:res_area} shows the volume specific interface area (area per volume) calculated from Eq.~\eqref{eq:interface_area} using the BIMM-2D parameter estimates $\widehat{w}_{12}$ and $\widehat{\sigma}_b$. The model results are compared to interface areas $A_{m}$ computed using \textit{marching cubes} \cite{Lewiner2003} in the binary artificial data before blurring and addition of noise. Note that while the marching cube algorithm provides an indication of the true interface area in the data, it does not necessarily provide the ground truth due to inherent errors related to vertex sampling frequency and sampling accuracy~\cite{Jorgensen2010}. The marching cubes results are, on average, \SI{3e-4}{vox^2/vox^3} larger than the model results, corresponding to 1-5 \%. 

%Discussion point
The issue with segmentation based quantification is that it can be challenging to assess the accuracy of the result without careful manual inspection. The ability to calculate an interface area or volume fraction estimate through the BIMM-2D provides us with a much needed support to segmentation based approaches. This is particularly important when studying the evolution of structural parameters in large time series. As seen above in Fig.~\ref{fig:Vs} and Fig.~\ref{fig:discussion_peaks_I1I2dev_} some methods can introduce a systematic bias that is dependent on material parameters (e.g. volume fractions or size distribution). If the goal is to study the evolution of these parameters it is indeed problematic that the evolving parameter itself systematically biases the result. Any estimation method including the BIMM will suffer from biases to some degree. However, having two independent and quite different approaches (segmentation and model fitting), that are biased in different ways, allows us to automatically detect growing discrepancies between the two approaches when performing bulk automatic analysis of tomograms.

\subsection{Microstructure Characterization of a Solid Oxide Fuel Cell Electrode}

\begin{figure}
\includegraphics[width=\mysize\columnwidth]{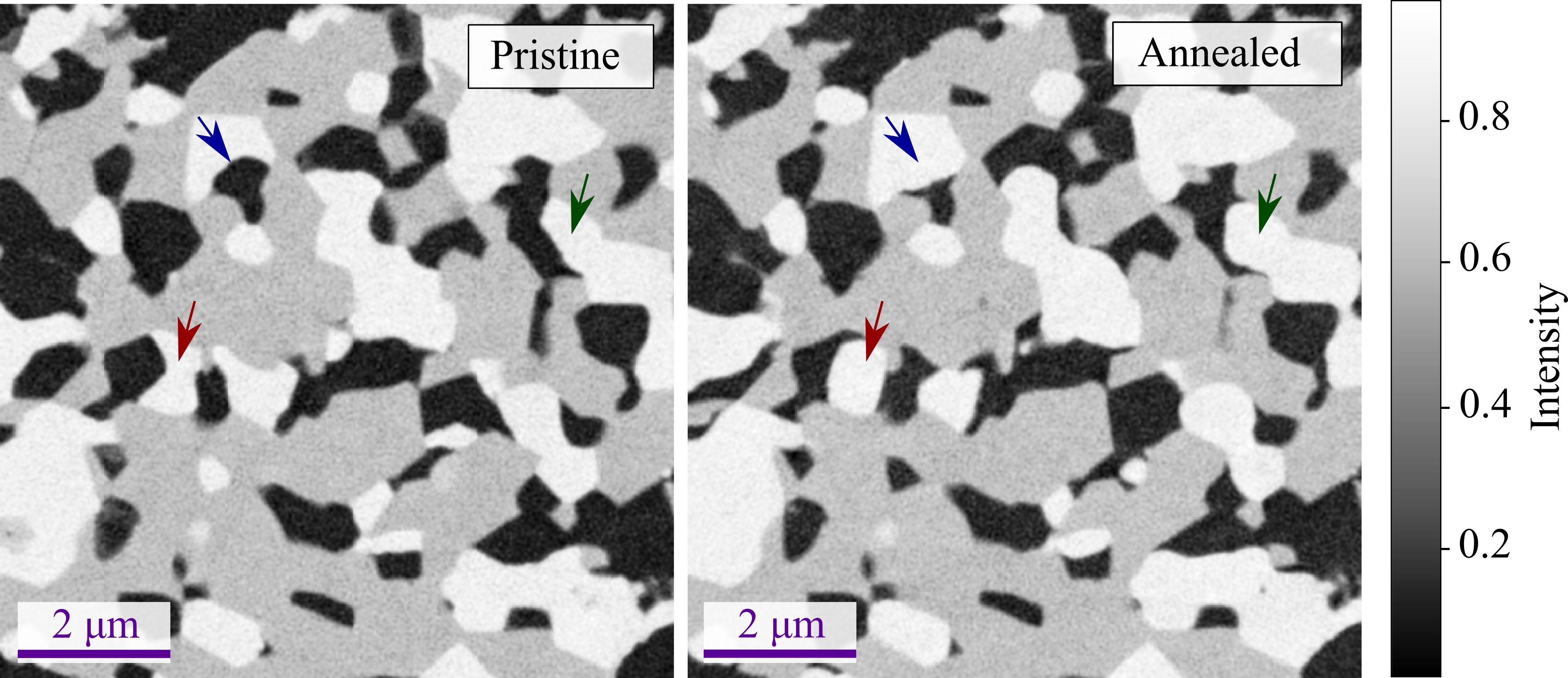}
\caption{Slices through 3-D datasets obtained by X-ray ptychography showing a fuel cell electrode before (left) and after (right) annealing~\cite{DeAngelis2018}. The three phases are pores (black), YSZ (grey) and nickel (white). The arrows indicate example locations where coarsening of nickel is visible, leading to a change in interface areas.  }\label{fig:real_data}
\end{figure}

%%% PRESENT THE SAMPLE
The proposed method for estimating volume fractions, resolution and interface areas was applied to  previously reported \cite{DeAngelis2018} experimentally obtained datasets with three phases. The sample is a solid oxide fuel cell (SOFC) electrode, a porous structure consisting of nickel (Ni) and yttrium-stabilized zirconia (YSZ). The datasets, shown in Fig.~\ref{fig:real_data}, are registered 3-D X-ray ptychography images of the sample before and after annealing at \SI{850}{\celsius} for 3 hours. By measuring volume fractions and interface areas in the two datasets, the change in microstructure due to annealing can be studied.

\begin{figure}
\includegraphics[width=\mysize\columnwidth]{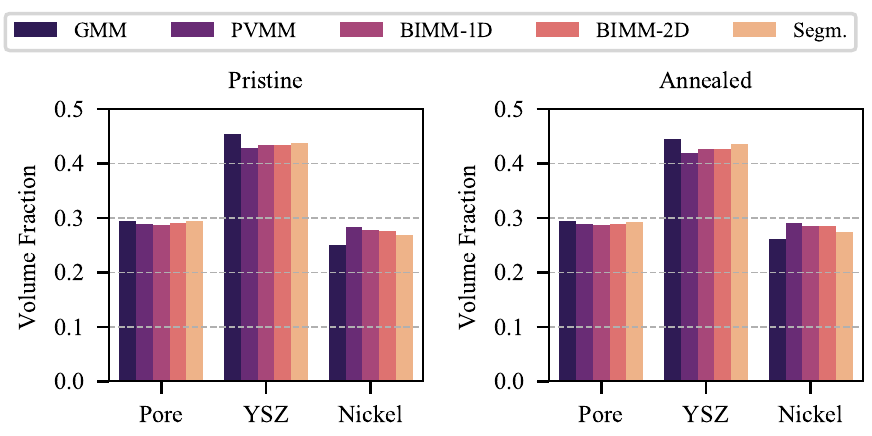}
\caption{Volume fractions for the fuel cell sample in the pristine and annealed state. The label ``Segm." refers to results obtained by counting voxels in segmented data. The other labels indicate which model was fitted to the raw data. Standard deviations are \SIrange{0.5}{1}{\percent}, highest for the GMM. }\label{fig:vf_pristine_treated}
\end{figure}

%% VOLUME FRACTIONS
Fig.~\ref{fig:vf_pristine_treated} shows the volume fractions obtained for the pristine and annealed dataset. Results reported by De Angelis et al.~\cite{DeAngelis2018} were obtained by counting voxels in data segmented by manual tuning of parameters for a 2D (intensity--gradient magnitude) histogram thresholding procedure. In the figure, the reported results (labeled "Segm.") are compared to results obtained with the GMM (component weights), the PVMM, and the BIMM (1D and 2D) (Eq.~\eqref{eq:volume_fractions}). All results are similar in value. Note that the reported volume fractions~\cite{DeAngelis2018} should not be seen as the ground truth; the true volume fractions of the analysed volumes are unknown.

%%% RESOLUTION
The blur parameter $\sigma_b$ of the BIMM-2D is estimated to $1.456 \pm 0.008$ and $1.452 \pm 0.007$ for the fuel cell sample in the pristine and the annealed state, respectively. Using Eq.~\eqref{eq:1090} and a voxel size of \SI{18.4}{\nano\meter} we get a resolution according to the 10 \%--90 \% criterion of $68.7 \pm 0.4 $ and $68.5 \pm 0.3$ nm, respectively. This corresponds well with the average value found when manually measuring three interface profiles in the pristine dataset, giving resolutions of 60 nm, 70 nm and 80 nm (see Fig.~\ref{fig:pristine_profiles}). 

%%% AREA  
Eq.~\eqref{eq:interface_area} allows us to estimate the area of specific interfaces in the three-phase data from the fitted BIMM-2D model parameters. Table~\ref{tab:area_pristine_treated_change} shows volume specific interface areas (area per volume) for the three interface types in the fuel cell sample, before and after annealing. Interface areas estimated using the BIMM-2D are compared to areas calculated by polygonization (meshing) of interfaces in segmented data \cite{Jorgensen2010}, as reported by De Angelis et al.~\cite{DeAngelis2018}. Although the two approaches compared here are very different, the obtained estimates are similar. In particular, we observe the same trend for both interface area calculation methods; after annealing, the area of the interface YSZ/Pore has increased, while the areas of the interfaces Pore/Ni and Ni/YSZ have decreased. Again, as the ground truth is not known, it is difficult to determine whether the statistical approach or the segmentation and meshing performs best. 

Looking closely at the pristine and annealed datasets in Fig.~\ref{fig:real_data}, we see some level of texture in the phase interiors. This is reflected in the voxel correlation estimated by the model parameter $\rho$, which is 0.47 and 0.42 for the pristine and annealed dataset, respectively ($\rho$ is constrained to the interval [-1,1]). As a similar texture is seen in all phases, including the pores (air), we can conclude that the texture is mainly a result of blurred noise rather than a physical texture in the materials. It indicates that the data has been blurred (filtered) to some degree during reconstruction or post-processing. In contrast, for the artificial data, no filter was applied and the voxel correlation was zero (Table~\ref{tab:data_gen_vs_model_est}).

\begin{table}
\caption{Interface areas [\SI{}{\micro \meter \squared / \micro \meter \cubed}],  pristine and annealed sample. Standard deviations for the BIMM-2D estimates are \SIrange{1.7}{2.7}{\percent}. }\label{tab:area_pristine_treated_change} 
%\begin{ruledtabular}
\begin{tabular}{cccc}
\toprule \\
       & Pore/YSZ & Pore/Ni & YSZ/Ni \\
      \bottomrule \\
      Pristine Sample \\
      \hline 
      \begin{tabular}{l} Segm.+mesh  \\ \textbf{BIMM-2D}  \end{tabular}   & \begin{tabular}{l} 1.18  \\ \textbf{1.09}  \end{tabular}  &\begin{tabular}{l}  0.64 \\ \textbf{0.64}  \end{tabular}  &\begin{tabular}{l}   0.97 \\ \textbf{0.81}  \end{tabular} \\
      \bottomrule \\
      Annealed Sample \\
      \hline 
       \begin{tabular}{l} Segm.+mesh  \\ \textbf{BIMM-2D}  \end{tabular}    & \begin{tabular}{l} 1.30  \\ \textbf{1.23}  \end{tabular}  &\begin{tabular}{l}  0.57\\ \textbf{0.60}  \end{tabular}  &\begin{tabular}{l}   0.83 \\ \textbf{0.69}  \end{tabular}  \\
      \bottomrule \\
      Relative Change \\
      \hline 
      \begin{tabular}{l} Segm.+mesh  \\ \textbf{BIMM-2D}  \end{tabular} & \begin{tabular}{l} 10 \% \\ \textbf{13 \%}  \end{tabular}  &\begin{tabular}{l}  -11 \%\\ \textbf{-7 \%}  \end{tabular}  &\begin{tabular}{l}  -14 \%  \\ \textbf{-15 \%} \end{tabular} \\
        \bottomrule 
\end{tabular}
%\end{ruledtabular}
\end{table}

\begin{figure*}
    \subfloat[]{\includegraphics[width=\mysizethird\textwidth]{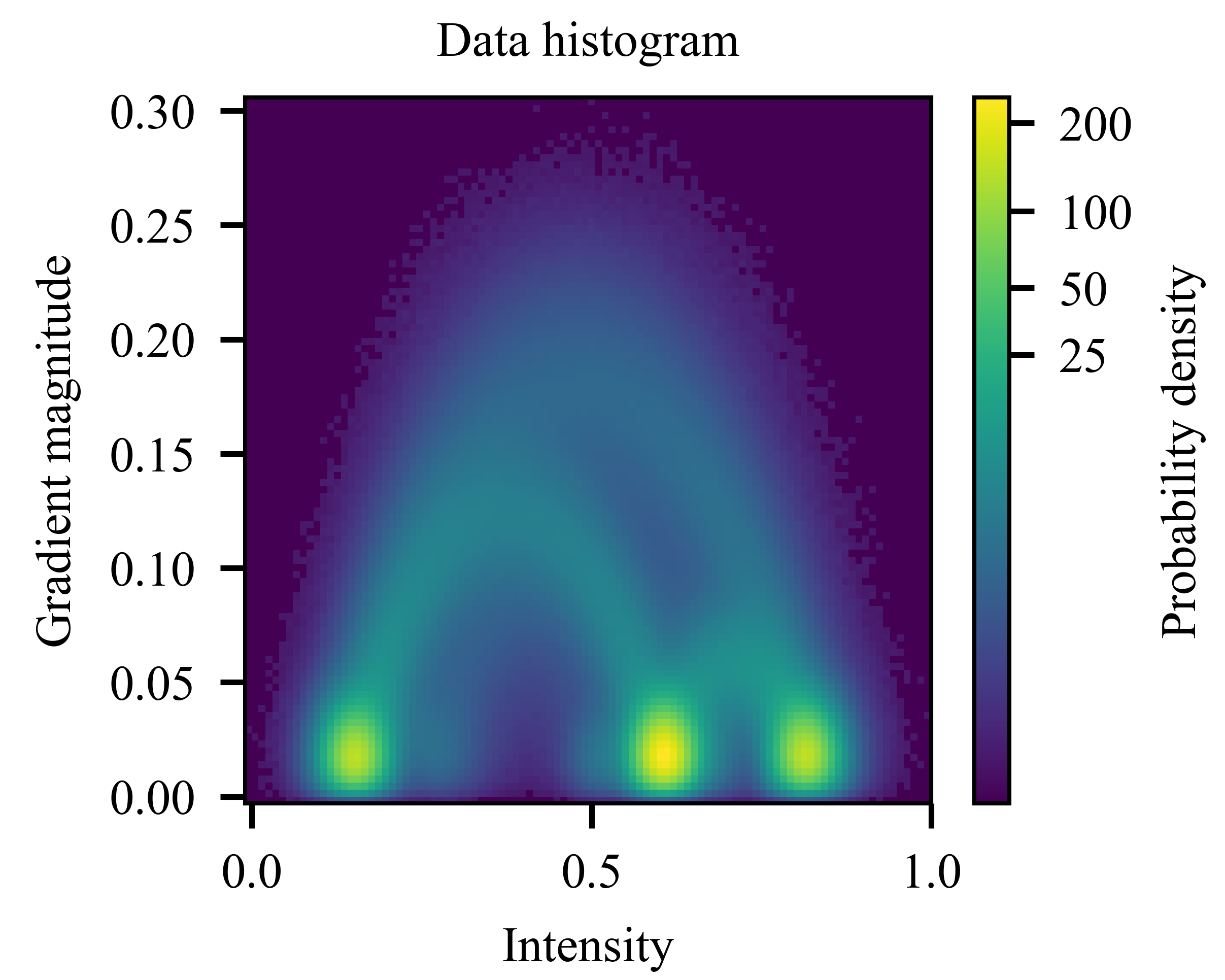}\label{fig:pristine_hist2D}}
    ~ %add desired spacing between images, e. g. ~, \quad, \qquad, \hfill etc. 
      %(or a blank line to force the subfigure onto a new line)
    \subfloat[]{\includegraphics[width=\mysizethird\textwidth]{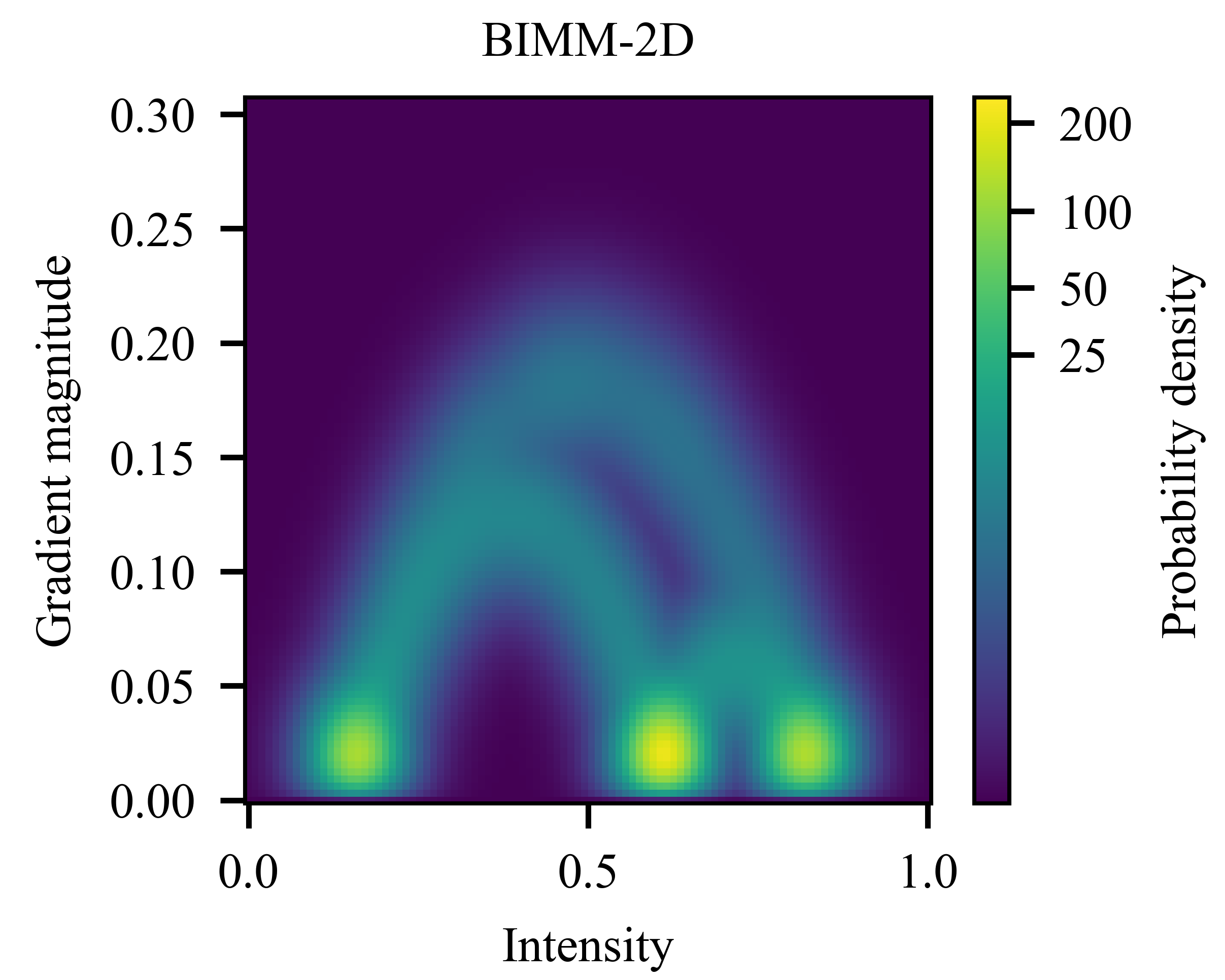}\label{fig:pristine_gridplot2D}}
    ~
    \subfloat[]{\includegraphics[width=\mysizethird\textwidth]{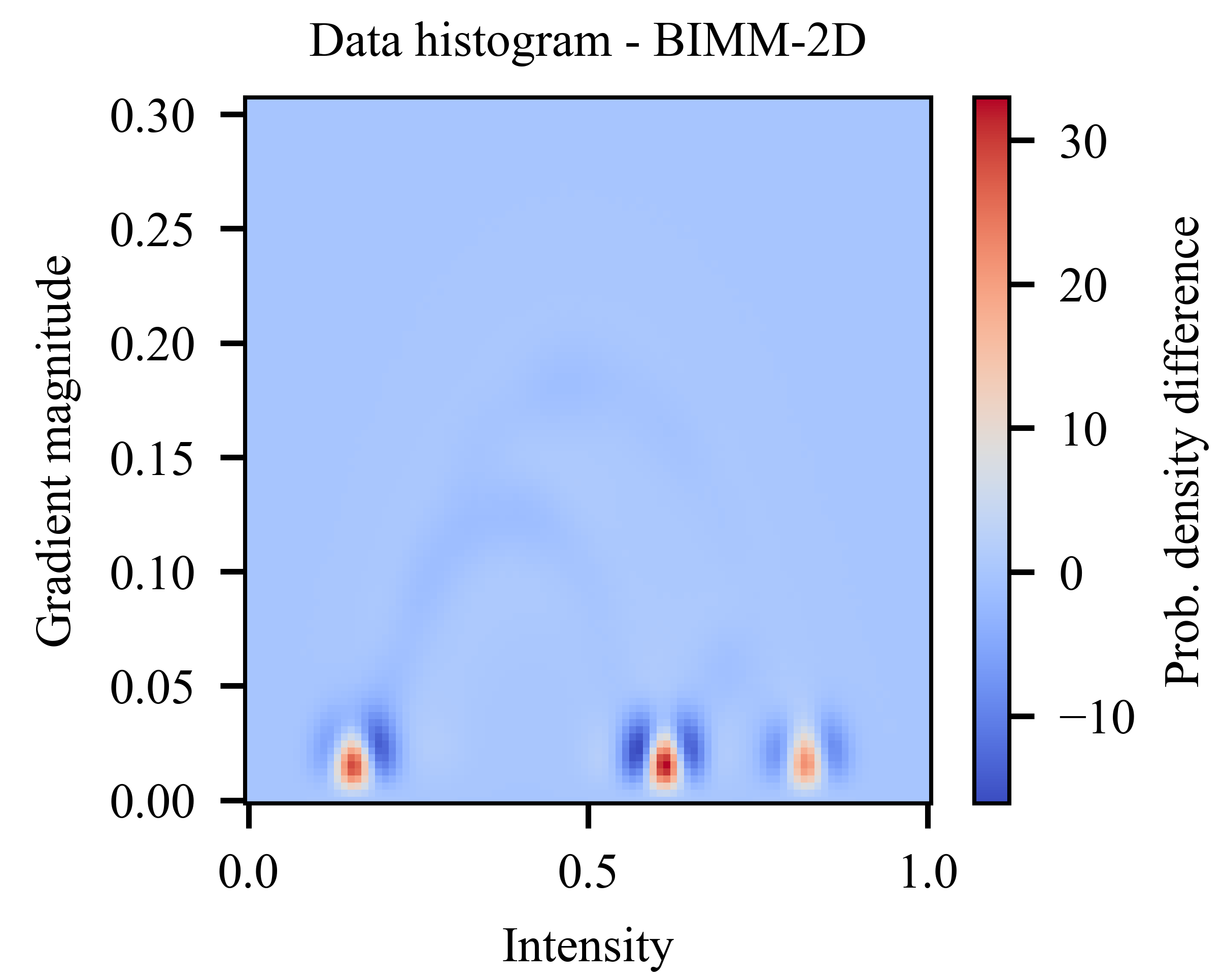}\label{fig:pristine_diff}}
    \caption{\textbf{(a)} 2D histogram of intensity and gradient magnitude values of the pristine fuel cell dataset. \textbf{(b)} Fitted BIMM-2D evaluated over the same region as the histogram. \textbf{(c)}  Model plot subtracted from the data histogram.  }\label{fig:model_vs_data_2D}
\end{figure*}

\begin{figure*}
    \subfloat[$L = -0.581$]{\includegraphics[width=\mysizethird\textwidth]{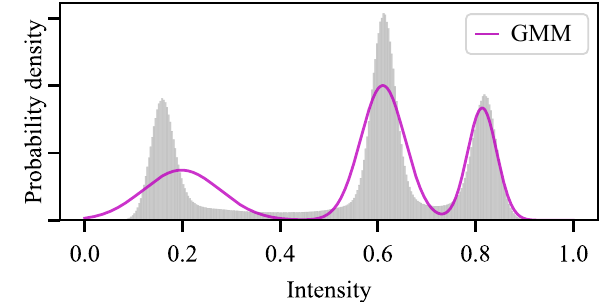}\label{fig:1Dhist_model_gmm}}
    ~ %add desired spacing between images, e. g. ~, \quad, \qquad, \hfill etc. 
      %(or a blank line to force the subfigure onto a new line)
    \subfloat[$L = -0.760$]{\includegraphics[width=\mysizethird\textwidth]{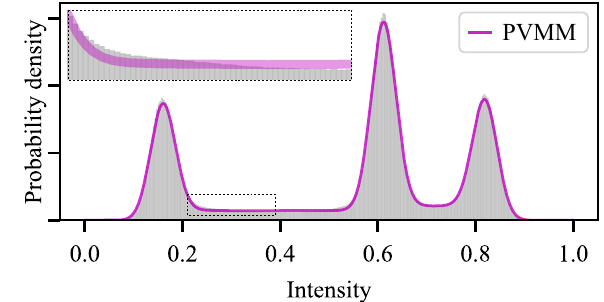}\label{fig:1Dhist_model_pvmm}}
    ~
    \subfloat[$L = -0.762$]{\includegraphics[width=\mysizethird\textwidth]{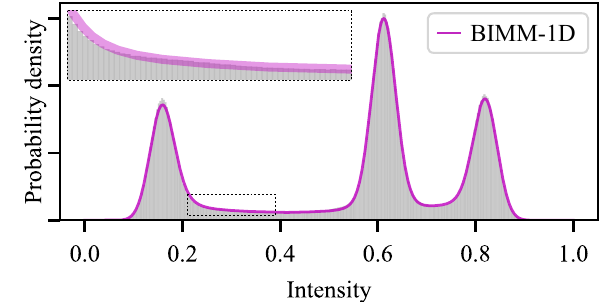}\label{fig:1Dhist_model_bimm1d}}
    \caption{Intensity histogram of the pristine fuel cell dataset (grey) with the PDF of fitted models (pink). A smaller value of the negative log-likelihood $L$ (Eq.~\eqref{eq:L}) indicates a better model fit.   }\label{fig:1Dhist_model}
\end{figure*}

%%% MODEL FIT
Visually assessing the fit of the model to the dataset is challenging, as seen in Fig.~\ref{fig:model_vs_data_2D} which provides a comparison between the 2D (intensity--gradient magnitude) histogram and the BIMM-2D. The negative log-likelihood $L$ from Eq.~\eqref{eq:L} can be used as a quantitative measure for the goodness of fit.  Fig.~\ref{fig:1Dhist_model} shows the fitted GMM, the PVMM and the BIMM-1D plotted with the normalized intensity histogram of the pristine dataset, with the $L$ values listed in the captions. Similar figures with individual model components plotted are found in the Supplementary Material, Fig.~S3. While the $L$ value alone is difficult to interpret, the relative value indicates that the BIMM-1D has a better fit than the GMM and the PVMM, agreeing with the qualitative visual impression. Insets show that interface voxels are better modelled by the BIMM-1D than by the PVMM. Having an objective measure of the model fit can be useful in many applications. For example, in an automatic workflow system processing a large number of datasets, such a measure can be useful to automatically flag data that need a manual check. The poor fit could be a result of severe image artifacts or large changes in imaging conditions, pointing out to the operator which datasets might be unsuitable for further processing. %~\ref{fig:1Dhist_model_components}

\subsection{Segmentation of 3-D Image Data with Multiple Phases }

\begin{figure}
    \includegraphics[width=\mysize\columnwidth]{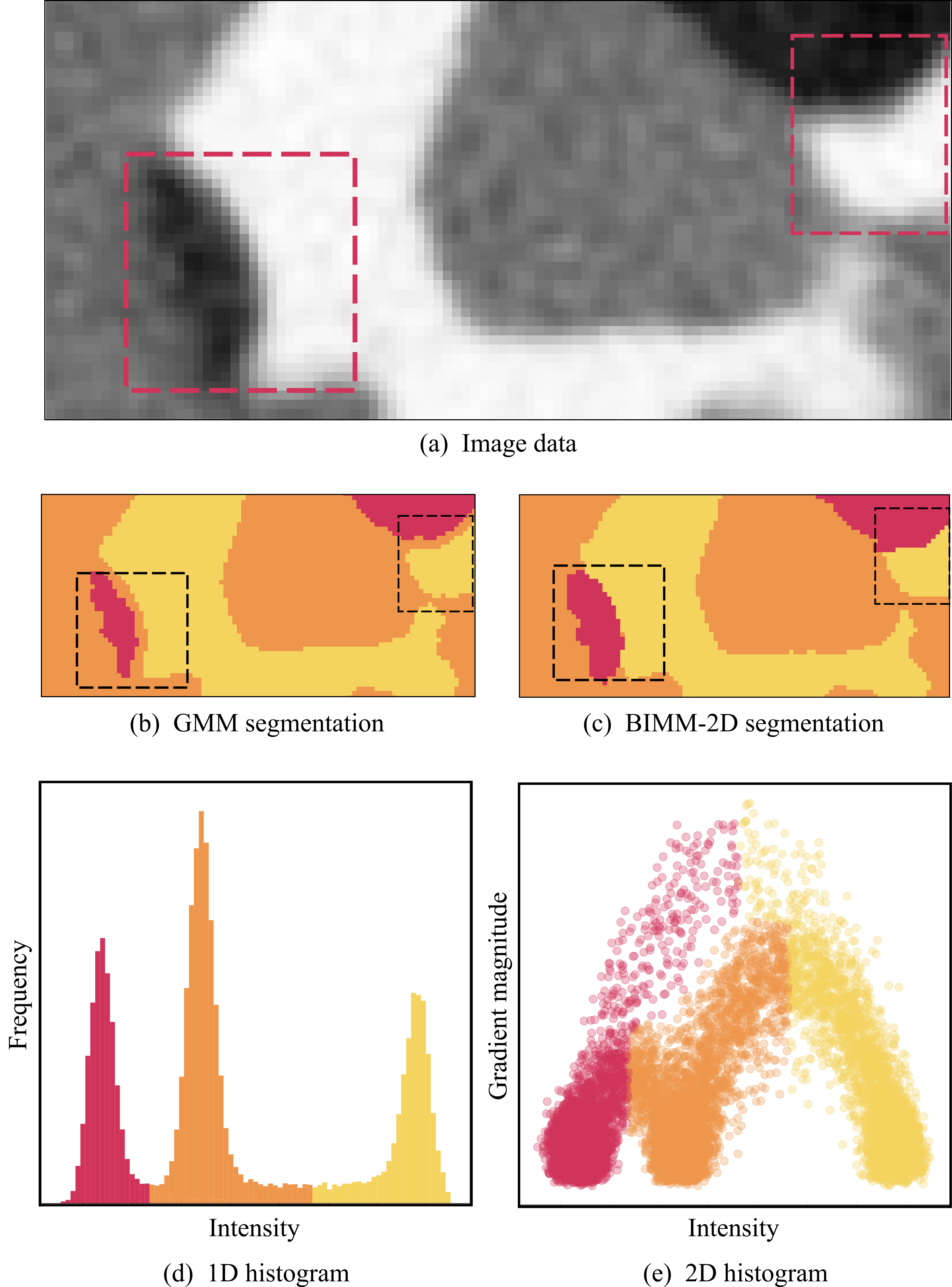}
    \caption{Comparison of maximum likelihood segmentations based on the GMM and the BIMM-2D. Our model performs better at interfaces between black and white voxels, indicated with stippled frames in (a), (b) and (c).  \textbf{(d)} 1D intensity histogram with colors according to labels in (b). \textbf{(e)} 2D intensity--gradient magnitude scatter plot with colors according to labels in (c).  }\label{fig:segm_with_hist}
\end{figure}

The derived model can be used as input for probabilistic segmentation methods. Here, we demonstrate the use of the model for a simple \textit{maximum likelihood segmentation} of the pristine fuel cell data (Fig.~\ref{fig:real_data}(a)), and compare the performance with the GMM. The result is shown in Fig.~\ref{fig:segm_with_hist}. While a quantitative assessment is challenging, the visual comparison is enough to spot that the BIMM-2D outperforms the GMM at interfaces between pores (black) and nickel (white). Overall, the result using the BIMM-2D looks reasonable. 

The maximum likelihood segmentation was performed by labeling each voxel according to the highest probability model component. The GMM contains 3 model components; one for each material phase (black, grey, white). We therefore get a segmentation into three phases directly (red, orange and yellow in the figure). For the BIMM-2D, we have 6 components resulting in 6 different labels; one for each phase and one for each interface type (black--grey, black--white,grey--white). The interface-labeled voxels are then re-assigned to one of the adjacent phases. As an example, a voxel labeled ``black--white" is classified as belonging to either the black or the white phase, while the grey phase is not considered. The re-labeling is done using a threshold at the midpoint between white and black, $(I_{white}+I_{black})/2$. 

For the GMM, only intensity information is considered and therefore this maximum probability segmentation corresponds to intensity thresholding as indicated in the 1D intensity histogram in Fig.~\ref{fig:segm_with_hist}(d). This mistakenly labels the interface between black and white voxels as belonging to the grey phase, as seen in Fig.~\ref{fig:segm_with_hist}(b). Very similar results were seen for the BIMM-1D (not shown in the figure). In contrast, the BIMM-2D also takes gradient information into account, resulting in better performance at interfaces as seen in Fig.~\ref{fig:segm_with_hist}(c). This segmentation corresponds to thresholding in the 2D intensity--gradient magnitude space as illustrated in Fig.~\ref{fig:segm_with_hist}(e), where the scatter plot is colored according to final labels.

\subsection{Perspectives on the Model for Segmentation}

Whilst in the example in the previous section, we made use of a very basic probabilistic segmentation method, it nevertheless illustrates the advantage of having gradient information and interface blurring included in the model, as it assists separating interfaces from phase interiors. One significant advantage of the BIMM-2D segmentation method is that it is not affected by operator bias, as is the case for other segmentation methods involving manual tuning of segmentation parameters or manual labeling of training data for supervised learning-based segmentation methods. Furthermore, the fitted model inherently provides a sanity-check on the segmentation results. Structure parameters measured in the segmented data can be compared with those estimated by the model. Although similar volume fractions and interface areas does not guarantee a high accuracy, it does to strengthen the confidence in the segmentation result. 

The described approach can be seen as a step towards a physics based segmentation, as it relies on parameters that describe the physical nature of the imaged sample (intensities, volume fractions, interface areas) and the imaging system (resolution, noise). The negative log-likelihood $L$ from Eq.~\eqref{eq:L} can be used as a quantitative measure of how well the model fits a specific dataset. If the model fits the data well, we have confidence that the segmentation is accurate and unbiased. If it does not, we know that possible results could be biased and a different approach to segmentation must be pursued. The ability to insert automatic analysis assertions into a tomographic measurement pipeline is becoming ever more important, as time series tomographic data becomes more prevalent and human assessment less viable.  

%%%%%%%%%%%%%%%%%%%%%%%%%%%%%%%%%%%%%%%%%%%%%%%%%%%%%%%%%%%%%%%%%%%%%%%%%%%%%%%%%%%%
\section{Properties of the Model Fitting}

\subsection{ Model Complexity and Computation Time} 

Compared to the GMM, the BIMM-1D contains additional model components representing the interfaces in the data. This increase in model complexity gives an increase in computation time, from 5 s (GMM) to 7 s (BIMM-1D) (data with 2 phases, batch size 50, 1000 MC samples, 2000 iterations) on an Intel® Core™ i7-7600U CPU at 2.80GHz. This is mainly due to the MC integration needed in the interface components. Going from BIMM-1D to BIMM-2D, the inclusion of gradient information further increases the complexity, and could in principle be expected to double the computation time. In practice, however, the current implementation of the modified Bessel function of the first kind, $I_{\frac{1}{2}}$ in Eq.~\eqref{eq:p_v_I}, is a bottleneck, resulting in a runtime of 4 min for the BIMM-2D (same data and optimization parameters). This is however not an inherent issue with the model, and will be worked out for future releases of the code.

\subsection{Optimization Parameters and Computation Time}\label{sec:discussion_parameters}

Doubling the batch size from 50 to 100 was seen to reduce the model parameter confidence intervals (result variance) by approximately 30 \%, but doubled the computation time. So, the choice of batch size is a trade-off between variance in the estimated model parameters and computation time. 

The number of samples used in the MC integration affects the computation time in a similar manner. The computation time increases linearly with the number of MC samples, so a low number is preferred. However, because the objective function (Eq.~\eqref{eq:L}) includes the log of an MC integral, using a low number of samples introduces a bias. Studies using 10, 100, 1000 and 10000 MC samples showed a significant reduction of bias when increasing the number of MC samples from 100 to 1000, but little effect when increasing further to 10000. We therefore conclude that 1000 MC samples is sufficient.

\subsection{Handling Large Datasets}  

As shown in Section~\ref{sec:results}, only a small amount of randomly sampled voxels is needed to fit the model. With batch size 50 and 2000 iterations, convergence was reached utilizing as little as 0.08 \% of the datasets containing 125 million datapoints (500 x 500 x 500 voxels). This demonstrates that the dataset size is no limitation when using a statistical approach such as the one presented in this paper. In contrast, quantification methods involving segmentation and subsequent geometrical measurements (e.g. meshing to find interface area) can be computationally more heavy and have a higher memory consumption for large datasets, as they involve treating each individual voxel. The small amount of data required to fit the model also means that a workflow involving several local measurements is feasible. Local properties estimated from smaller subvolumes can thus be compared across the full sample.

%%%%%%%%%%%%%%%%%%%%%%%%%%%%%%%%%%%%%%%%%%%%%%%%%%%%%%%%%%%%%%%%%%%%%%%%%%%%%
\section{Conclusions}\label{sec:conclusions}

In this paper, we have derived a model for the distribution of intensity and gradient magnitude values in 3D X-ray tomography data. The model is based on a Gaussian approximation to the imaging point spread function (PSF) and assumes additive Gaussian noise with some correlation between voxels. The model can be seen as an extended Gaussian mixture model that takes blurred interfaces between material phases into account, and is therefore named the blurred interface mixture model (BIMM). The inclusion of gradient data in the model allows several additional physical parameters to be estimated directly, such as image resolution and the area of interfaces between phases. 

The BIMM outperforms existing statistical models in materials quantification. Compared to ground truth from artificial data, the fitted model parameters for phase intensities, noise level and interface blur level were seen to agree well. The BIMM provides more accurate estimates of material volume fractions compared to the Gaussian mixture model (GMM), and for phase intensity estimation, the BIMM was more accurate than both the GMM and the partial volume mixture model (PVMM), especially for less abundant material phases. Interface areas estimated by the BIMM-2D were seen to be similar to values calculated using marching cubes in the binary ground truth datasets.

We have demonstrated the applicability of the model for quantification and segmentation of experimentally obtained 3D X-ray CT data. As ground truth is not available for experimental data, results obtained by the BIMM-2D are compared to results obtained by first segmenting the data, then counting voxels to find volume fractions and fitting a mesh to find surface areas. The results of the two different approaches were found to be similar. Although assessing the accuracy of the result is challenging for any method, the ability to calculate interface areas or volume fractions using the BIMM provides us with a much needed independent support to segmentation based approaches. For segmentation of data with three phases, the BIMM-2D outperformed the GMM by avoiding critical misclassifications at interfaces.

%%%%%%%%%%%%%%%%%%%%%%%%%%%%%%%%%%%%%%%%%%%%%%%%%%%%%%%%%%%%%%%%%%%%%%%%%%%%%%%%%%%

\section*{Acknowledgments}
This project has received funding from the European Union’s Horizon 2020 research and innovation programme under the Marie Sk\l odowska-Curie grant agreement No 765604.  %Skłodowska

\section*{Declaration of competing interest}
The authors declare that they have no known competing financial interests or personal relationships that could have appeared to influence the work reported in this paper.

\section*{Data Availability}
%The raw data required to reproduce these findings are available to download from [INSERT PERMANENT WEB LINK(s)]. The processed data required to reproduce these findings are available to download from [INSERT PERMANENT WEB LINK(s)].
The developed code is available at \url{https://github.com/elobre/bimm}. Examples of usage are included, showing the BIMM used for quantification and segmentation of artificial and experimental data.

The code for generating the artificial data required to reproduce these findings are available to download from the same link. The experimental data required to reproduce these findings are available to download from DOI: 10.5281/zenodo.1040274 \cite{DeAngelis2018Dataset}.

%%%%%%%%%%%%%%%%%%%%%%%%%%%%%%%%%%%%%%%%%%%%%%%%%%%%%%%%%%%%%%%%%%%%%%%%%%%%%%%%%%

\appendix

\section{Details on the Model Derivation}\label{app:model_derivation}

\subsection{Derivation of \texorpdfstring{$G(I)$}{\textit{G(I)}}}

For an interface between materials of intensity $I_i$ and $I_j$, oriented so that the normal points along the $x$-axis, the expression for the intensity profile along the interface normal from Eq.~\eqref{eq:I_x}, and repeated here for convenience, is
\begin{linenomath}\begin{equation}\label{eq:I_x_app}
I(x, y, z) = I_i + (I_j-I_i)  \cdot \frac{1}{2} \Big[1+\text{erf} \Big(\frac{x}{\sigma_b \sqrt{2}}\Big)\Big].
\end{equation}\end{linenomath}
The \textit{gradient magnitude} $ \mid \nabla I(x,y,z) \mid$ is defined as the root of the squared sum of gradient components along the $x$, $y$ and $z$ axis,
\begin{linenomath}\begin{equation}
    \begin{split}
        & \mid \nabla I(x,y,z) \mid \\
        &= \sqrt{ \Big(\frac{\partial }{\partial x} I(x,y,z)\Big)^2 + \Big(\frac{\partial}{\partial y} I(x,y,z)\Big)^2 + \Big(\frac{\partial }{\partial z} I(x,y,z)\Big)^2} .
    \end{split}
\end{equation}\end{linenomath}
In this case, the gradient components along the $y$ and $z$ axis are zero, so the gradient magnitude is equivalent to the derivative of Eq.~\eqref{eq:I_x_app} with respect to $x$, 
\begin{linenomath}\begin{equation}\label{eq:nabla_I}
\mid \nabla I(x,y,z) \mid = \left\vert \frac{\text{d}}{\text{d}x} I(x, y, z) \right\vert =  \frac{\mid I_j-I_i \mid}{ \sigma_b\sqrt{2\pi}}  \text{exp}\Big(  - \frac{x^2}{2\sigma_b^2} \Big). 
\end{equation}\end{linenomath}
In order to express the gradient magnitude in terms of $I$, we invert Eq.~\eqref{eq:I_x_app},
\begin{linenomath}\begin{equation}\label{eq:x_I} 
x(I) = \sigma_b\sqrt{2} \, \text{erf}^{-1}\left(2\frac{I-I_i}{I_j-I_i}-1\right).
\end{equation}\end{linenomath}
Substituting for $x$ in Eq.~\eqref{eq:nabla_I}, we get the gradient magnitude expressed in terms of intensity, 
\begin{linenomath}\begin{equation}\label{eq:G_I_app} 
\begin{split}
G(I) &= \left\lvert \nabla I(I)  \right\rvert \\
&= \frac{ \mid I_j-I_i \mid }{\sigma_b \sqrt{2\pi} } \text{exp}\Big(-\Big[ \text{erf}^{-1}\Big(2\frac{I-I_i}{I_j-I_i}-1\Big)\Big]^2 \Big).
\end{split}
\end{equation}\end{linenomath}
Note that, because this expression is independent of the choice of coordinate system, it is valid for interfaces at all orientations.

\subsection{Approximating \texorpdfstring{$G(I)$}{\textit{G(I)}} through Central Differences}\label{app:central_diff_num}

The components of the gradient magnitude $ \mid \nabla I(x,y,z) \mid $ is given by
\begin{linenomath}\begin{equation}\label{eq:G_z_lim}
\frac{\partial}{\partial x} I(x,y,z)=  \lim_{h \to 0} \frac{ I(x+h) - I(x-h)}{2h}.
\end{equation}\end{linenomath}
For central differences in discrete image data, the smallest possible $h$ without interpolation\footnote{Interpolation would introduce an unwanted smoothing, altering the noise distribution we seek to model.} is the width of one voxel, $h=1$. The gradient component is therefore approximated using the intensities of the neighbouring voxels, 
\begin{linenomath}\begin{equation}\label{eq:dIx}
\frac{\partial}{\partial x} I(x,y,z) \approx \frac{1}{2} \Big[ I(x-1,y,z) - I(x+1,y,z) \Big]. 
\end{equation}\end{linenomath} 
The central differences approximation for the gradient magnitude in 3-D is therefore as follows, 
\begin{linenomath}\begin{equation}\label{eq:G_I_approx}
\begin{split}
G(I)  &\approx \left\{  \left(\frac{1}{2}\left[ I(x-1,y,z) - I(x+1,y,z) \right]\right)^2  \right. \\ 
&+ \left. \left(\frac{1}{2}\left[ I(x,y-1,z) - I(x,y+1,z) \right]\right)^2  \right. \\ 
&+ \left.  \left(\frac{1}{2}\left[ I(x,y,z-1) - I(x,y,z+1) \right]\right)^2 \right\} ^{\frac{1}{2}}
\end{split}
\end{equation}\end{linenomath}

\subsection{The Statistical Distribution of the Gradient Magnitude Data \texorpdfstring{$\vv$}{\textbf{\textit{v}}}}\label{app:dist_v}

In this section, derive the probability density function (PDF) of gradient magnitudes $\vv$, and see that it can be expressed in terms of $G(I)$ through the central differences approximation in Eq.~\eqref{eq:G_I_approx}.

Using central differences, the gradient magnitude $v$ for the voxel at coordinates $(x,y,z)$ is calculated from the intensities of the six nearest-neighbouring voxels,

\begin{linenomath}\begin{equation}\label{eq:Y}
\begin{split}
v &=\sqrt{ v_x^2 + v_y^2 + v_z^2} =    \left\{ \left(\frac{1}{2}\left[u_{x-1}-u_{x+1}\right]\right)^2 \right. \\
&+ \left. \left(\frac{1}{2}\left[u_{y-1}-u_{y+1}\right]\right)^2 + \left(\frac{1}{2}\left[u_{z-1}-u_{z+1}\right]\right)^2   \right\}^{\frac{1}{2}}.
\end{split}
\end{equation}\end{linenomath}
With this notation, $u_{x-1}$ and $u_{x+1}$ are the intensities of the neighbouring voxels in the $x$-direction, etc. As we assume additive Gaussian noise, the voxel intensity $u$ is normally distributed,
\begin{linenomath}\begin{equation*}
u \sim N(I(x, y, z), \sigma_n^2).
\end{equation*}\end{linenomath}
Therefore, each gradient component $v_j$ with $j=x,y,z$ also follows a normal distribution, $v_j \sim N(\mu_{v_j}, \sigma^2_{v_j})$. The gradient magnitude $v$ is thus the root of the sum of squared normally distributed variables. To arrive at the PDF of $v$, we first find the expected values and variance of the gradient components $v_j$,
\begin{linenomath}\begin{equation*}
\begin{split}
\sigma_{v_j}^2 &= \text{Var}[ v_j ]  =  \text{Var}\left[  \frac{1}{2} (u_{j-1} - u_{j+1}) \right] \\
&= \left( \frac{1}{2} \right)^2 \left( \text{Var}[u_{j-1}] + \text{Var}[u_{j+1}] - 2\, \text{Cov}[u_{j-1}, u_{j+1}]\right),
\end{split} 
\end{equation*}\end{linenomath}
so
\begin{linenomath}\begin{equation}\label{eq:component_var}
\sigma_{v_x}^2 =\sigma_{v_y}^2 =\sigma_{v_z}^2 = \frac{1}{4} ( \sigma_n^2 + \sigma_n^2 - 2 \sigma_n^2 \rho) = \frac{\sigma_n^2}{2}(1-\rho),
\end{equation}\end{linenomath} 
where $\rho$ denotes the correlation between $u_{j-1}$ and $u_{j+1}$, and
\begin{linenomath}\begin{equation*}
 \mu_{v_j} = \text{E} [ v_j ] =  \text{E}\left[  \frac{1}{2} (u_{j-1} - u_{j+1}) \right] = \frac{1}{2} \left( \text{E}[u_{j-1}] - \text{E}[u_{j+1}]\right), 
 \end{equation*}\end{linenomath}
so
\begin{subequations}\label{eq:grad_component_means}
\begin{eqnarray}
%    \begin{split}
        \mu_{v_x} &= \frac{1}{2} \left[ I(x-1,y,z) - I(x+1,y,z) \right],  \\ 
        \mu_{v_y} &= \frac{1}{2} \left[ I(x,y-1,z) - I(x,y+1,z) \right],  \\ 
        \mu_{v_z} &= \frac{1}{2} \left[ I(x,y,z-1) - I(x,y,z+1) \right]. 
 %   \end{split}
\end{eqnarray}
\end{subequations}
Now, with the same variance for all gradient components (Eq.~\eqref{eq:component_var}), the gradient magnitude $v$ has a non-central chi distribution with three degrees of freedom, scaled by the root of the component variance,
\begin{linenomath}\begin{equation*} v \sim \text{NC}\chi_{k=3}(\lambda) \cdot  \frac{\sigma_n\sqrt{1-\rho}}{\sqrt{2}}
\end{equation*}\end{linenomath}
with the non-centrality parameter $\lambda$ defined as
\begin{linenomath}\begin{equation*} 
\lambda = \sqrt{ \sum\limits_{j=x,y,z} \left( \frac{\mu_{v_j}}{\sigma_{v_j}} \right)^2 } = \frac{\sqrt{2}} {\sigma_n\sqrt{1-\rho}} \sqrt{ \mu_{v_x}^2+\mu_{v_y}^2+\mu_{v_z}^2} \end{equation*}\end{linenomath}
Then, with Eq.~\eqref{eq:G_I_approx} and Eqs.~\eqref{eq:grad_component_means}, $\lambda$ can be approximated using the analytical expression for the gradient, $G(I)$:
\begin{linenomath}\begin{equation}\label{eq:lambda_app}
    \lambda  \approx \frac{\sqrt{2}}{\sigma_n\sqrt{1-\rho}} G(I).
\end{equation}\end{linenomath}
From this, we can find the PDF of $v$. We introduce the helper variable $C \sim \text{NC}\chi_{k=3}(\lambda) $ such that $v= C \,\frac{ \sigma_n\sqrt{1-\rho}}{\sqrt{2}}$. The PDF of $C$ is given by
\begin{linenomath}\begin{equation*}
p_C(c) =   \frac{c^3 \lambda}{(\lambda c)^{3/2}} \text{exp}\left(-\frac{c^2 + \lambda^2}{ 2} \right) I_{1/2}(\lambda c).
\end{equation*}\end{linenomath}
Here, $I_{\frac{1}{2}}$ is the modified Bessel function of first kind. With the helper function $c(v) = v \frac{\sqrt{2}}{\sigma_n\sqrt{1-\rho}}$, the PDF of $v$ is found through a change of variables, 
\begin{linenomath}\begin{equation*}
p_V(v) = p_C(c(v))\,  \lvert \partial_v \, c(v) \rvert = p_C \left( \frac{v \sqrt{2}}{\sigma_n\sqrt{1-\rho}}\right)   \frac{\sqrt{2}}{\sigma_n\sqrt{1-\rho}}.
\end{equation*}\end{linenomath}
With $\lambda$ from Eq.~\eqref{eq:lambda_app} we get that the PDF of $v$ depends on $I$ through $G(I)$. Omitting the subscript $V$ to adhere with the notation used in Sec.~\ref{sec:method}, we arrive at
\begin{linenomath}\begin{equation}\label{eq:p_v_I_app} 
\begin{split}
p( v \mid I) &= \frac{2}{\sigma_n^2(1-\rho)} \sqrt{\frac{ \vv^3}{G(I) }}  \text{exp}\left(-\frac{\vv^2+G(I)^2}{\sigma_n^2(1-\rho)} \right) \\
& I_{\frac{1}{2}}\left(\frac{2 \vv G(I)}{\sigma_n^2(1-\rho)}\right).
\end{split}
\end{equation}\end{linenomath}

\subsection{Model fitting using Mini-Batch Gradient Descent with Adadelta}\label{app:gradient_descent}

Here, we provide more details on how we use maximum likelihood estimation (MLE) and gradient descent to find the optimal fit of our model $p(\vu,\vv)$ to the dataset. This method finds the model parameters $\vtheta_{max}$ that minimizes the difference between the model distribution and the data distribution, or in other words, maximizes the likelihood of the data. The likelihood function for a dataset $(\vu, \vv)$ is the joint PDF, $\prod_{m=1}^M  p(u_m,v_m)$, evaluated for all observations $u_m, v_m$ in the dataset. For the sake of numerical stability, it is common to instead minimize the negative log-likelihood, also called \textit{loss},
\begin{linenomath}\begin{equation}\label{eq:L_app}
 L(\vtheta)=  -\log \prod_{m=1}^M  p(u_m,v_m)= -\sum_{m=1}^M \log p(u_m,v_m).
\end{equation}\end{linenomath}

Gradient descent is one of the most common algorithms for optimization~\cite{Ruder2016}. In each iteration, the parameters $\vtheta$ are updated by $\Delta \vtheta$ in the direction of the negative gradient of the objective function $L(\vtheta)$:
\begin{linenomath}\begin{equation}
\begin{aligned}
    \Delta \vtheta_t &= - \eta \nabla_{\vtheta} L(\vtheta_t),\\
    \vtheta_{t+1} &= \vtheta_t +  \Delta \vtheta_t.
\end{aligned}
\end{equation}\end{linenomath}
In other words, for each iteration, we take one step in the direction of the slope of the surface represented by the objective function, towards a (local) minimum. The learning rate $\eta$ determines the step size. To set this, we use Adadelta, a method for per-parameter adaptive learning rates which requires no manual tuning and is insensitive to hyperparameters~\cite{Zeiler2012}. 

For large datasets, evaluating $ L(\vtheta_t)$ to compute the gradients $\nabla_{\vtheta} L(\vtheta_t)$ becomes computationally expensive, even unfeasible for datasets that are too big to fit in memory. A common approach is therefore to instead calculate an \textit{estimate} of the gradients using a randomly selected subset of the data (called \textit{mini-batch}) in each parameter update. This leads to a significant speed-up at the cost of a higher variance and lower convergence rate.

\subsubsection{PyTorch and Automatic Differentiation}

The optimization routine is implemented in PyTorch~\cite{Paszke2019}, an open-source Python library for machine learning which provides an easy-to-use framework for optimization using \textit{automatic differentiation}~\cite{Baydin2018}. This technique is based on computational graphs, keeping track of all operations performed on each parameter when calculating the loss; the so-called \textit{forward pass}. To find the derivative of the loss w.r.t. any parameter, $\nabla_{\vtheta} L$, the chain rule is applied at each node of the computational graph as one propagates backwards through the graph (\textit{back-propagation}). This way, you avoid the laborious and error-prone attempts on deriving and implementing the analytical derivatives. Compared to numerical differentiation (e.g. finite difference approximations), which is the other alternative, automatic differentiation is more accurate, more numerically stable and less computationally heavy. 

\subsection{Monte Carlo Gradient Estimation}\label{app:MCgrad}

In Section~\ref{sec:p_u_v}, we saw that we need Monte Carlo integration~\cite{Metropolis1949} to find the interface component of the model $p(\vu,\vv)$. The Monte Carlo estimate of the integral in Eq.~\eqref{eq:p_ij_u_v} is as follows,

\begin{linenomath}\begin{equation}\label{eq:p_ij_u_v_MC}
\begin{split}
p_{ij}(\vu,\vv) &= \int p(\vu \mid I)\, p_{ij}(\vv \mid I)\, p_{ij}(I)\text{d} I \\
&\approx \frac{1}{N} \sum_n^{N} p(\vu \mid I_n ) \, p_{ij}(\vv \mid I_n  ) \,\,\, \text{with} \,\,  I_n \sim p_{ij}(I ) .
\end{split}
\end{equation}\end{linenomath}
For the optimization, we need to calculate gradients of the loss with respect to the parameters $\vtheta$, where the stochastic component above is included. Now, a problem arises because the distribution we draw MC samples from, $ p_{ij}(I)$, is defined by these parameters $\vtheta$ (Eq.~\eqref{eq:p_ij_I}). Computing stochastic gradients is a well-known problem in machine learning and across statistical sciences; the problem of Monte Carlo gradient estimation~\cite{Mohamed2019}. We need to use the \textit{reparametrization trick}~\cite{Kingma2014} to instead draw MC samples from a distribution independent of the model parameters $\vtheta$. We therefore introduce a new variable $\epsilon$ which is standard uniformly distributed, 
\begin{linenomath}\begin{equation*}
\epsilon \sim U([0, 1]), \, \, \, \, p(\epsilon) = 1.
\end{equation*}\end{linenomath}
We now want to rewrite the integral in Eq.~\eqref{eq:p_ij_u_v_MC} so that instead of sampling from  $ p_{ij}(I)$ we can sample from $p(\epsilon)$ which is independent of $\theta$. Through change of variables, we have

\begin{linenomath}\begin{equation*}
x = d_s \sigma_b (2 \epsilon - 1),  \,\,\,\, p(x) = p(\epsilon ) \left\vert \frac{ \partial x }{\partial \epsilon } \right\vert^{-1}  .\end{equation*}\end{linenomath}
Combining this with Eq.~\eqref{eq:p_ij_I}, we have 
\begin{linenomath}\begin{equation*}
p_{ij}(I) = p(\epsilon ) \left\vert \frac{ \partial x }{\partial \epsilon } \right\vert^{-1} \left\vert \frac{ \partial I }{\partial x} \right\vert^{-1} =   p(\epsilon ) \left\vert \frac{ \partial I }{\partial \epsilon } \right\vert^{-1},
\end{equation*}\end{linenomath}
so 
\begin{linenomath}\begin{equation*}
p(\epsilon) = p_{ij} (I ) \left\vert  \frac{ \partial I}{ \partial \epsilon} \right\vert. 
\end{equation*}\end{linenomath}
Using this, we can rewrite Eq.~\eqref{eq:p_ij_u_v_MC} as follows,
\begin{linenomath}\begin{equation}
    \begin{split}
        p_{ij}(\vu,\vv) &= \int_{I_{min}}^{I_{max}} p(\vu \mid I)\, p(\vv \mid I)\, p_{ij}(I) \text{d} I \\
        &= \int_{0}^{1} p(\vu \mid I(\epsilon))\, p(\vv \mid I(\epsilon))\, p_{ij}(I(\epsilon)) \left\vert  \frac{ \partial I}{ \partial \epsilon} \right\vert \text{d} \epsilon \\
        &= \int_{0}^{1} p(\vu \mid I(\epsilon))\, p(\vv \mid I(\epsilon))\, p(\epsilon)  \text{d} \epsilon  \\
        &\approx \frac{1}{N} \sum_n^{N} p(\vu \mid I(\epsilon_n) ) \, p(\vv \mid I(\epsilon_n)  ) \,\,\, \text{with} \,  \epsilon_n \sim p(\epsilon) 
    \end{split}
\end{equation}\end{linenomath} 
Here, $I(\epsilon)$ is the interface profile from Eq.~\eqref{eq:I_x} expressed in terms of $\epsilon$. This reparametrization of the MC estimate ensures correct gradient estimates in the back-propagation.

\end{document}

\begin{frontmatter}

\title{\normalsize A Physical Model for Microstructural Characterization and Segmentation of 3D Tomography Data\\ \Large---Supplementary Material---}

\author[address_energy]{Elise O. Brenne\corref{mycorrespondingauthor}}
\cortext[mycorrespondingauthor]{Corresponding author}
\ead{elbre@dtu.dk}

\author[address_compute]{Vedrana A. Dahl }
\author[address_energy]{Peter S. J\o rgensen}

\address[address_energy]{Department of Energy Conversion and Storage, Technical University of Denmark, Fysikvej, 2800 Kgs. Lyngby, Denmark}
\address[address_compute]{Department of Applied Mathematics and Computer Science, Technical University of Denmark, Richard Petersens Plads, 2800 Kgs. Lyngby, Denmark}

\end{frontmatter}

\section{Data Generation and Model Fitting Results}
\noindent
\begin{figure}[!htbp]
\centering
\includegraphics[width=0.99\textwidth]{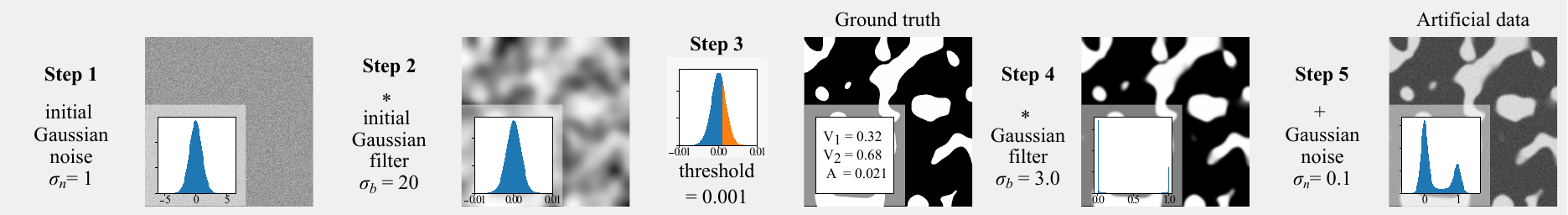}
\includegraphics[width=0.99\textwidth]{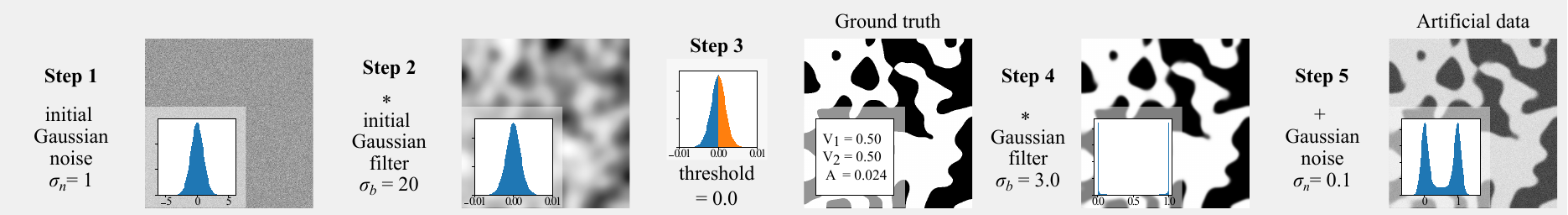}
\caption{\textbf{Artificial data generation.} Slices through the volume at each step of the data generation is shown for two of the artificial datasets. The intensity threshold in Step 3 determines the volume fractions. Ground truth volume fractions $V_1$ and $V_2$ and volume specific interface area $A$ (unit: \SI{}{vox \squared / vox \cubed}) are measured in the binarized volume after Step 3.  }\label{fig:art_data_gen}
\end{figure}
\begin{figure}
\centering
\hbox{
\includegraphics[width=0.3\textwidth]{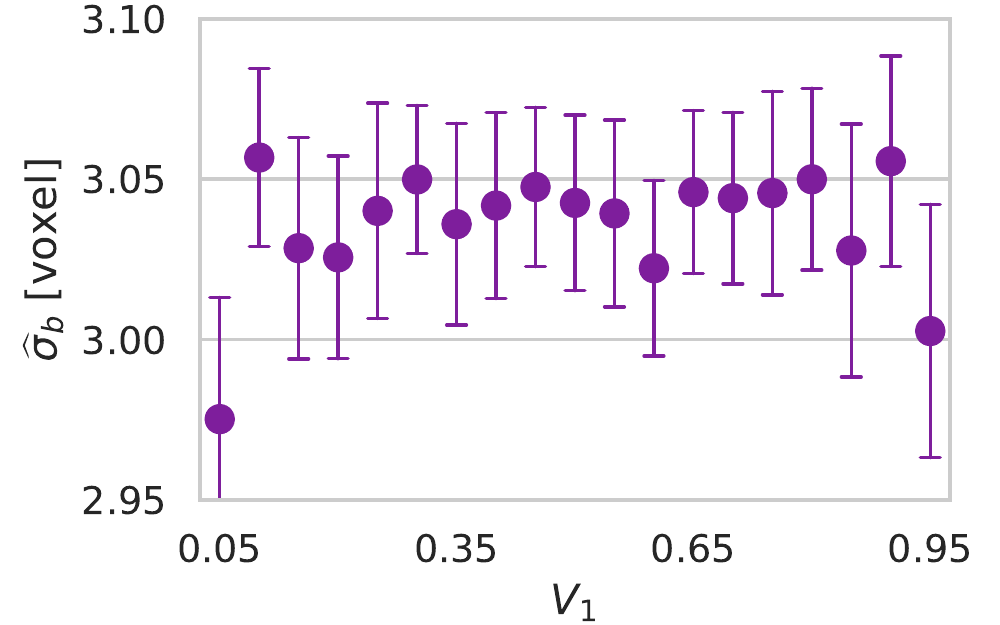}
\includegraphics[width=0.3\textwidth]{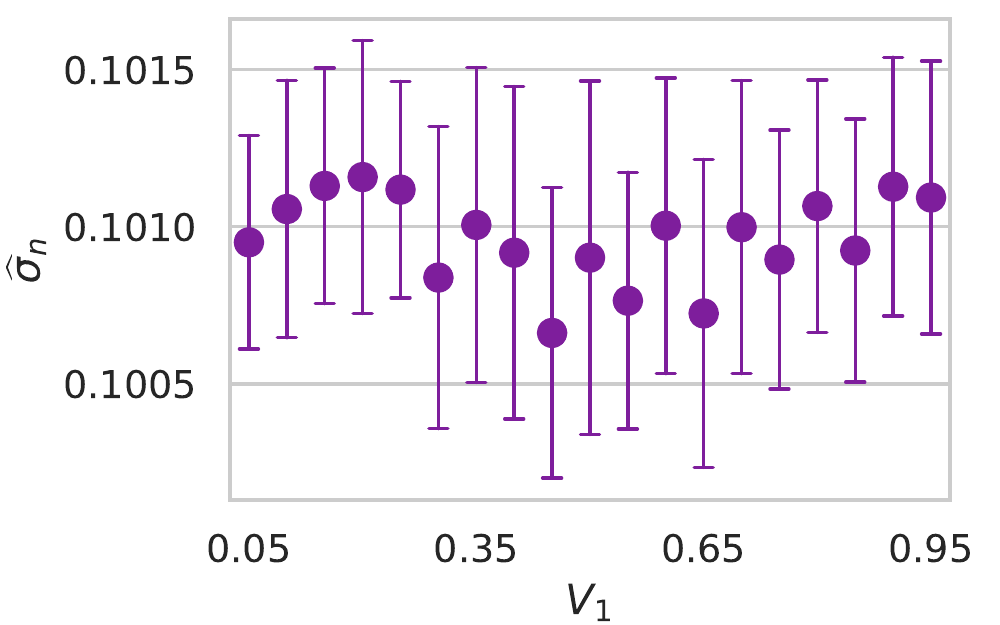}
\includegraphics[width=0.3\textwidth]{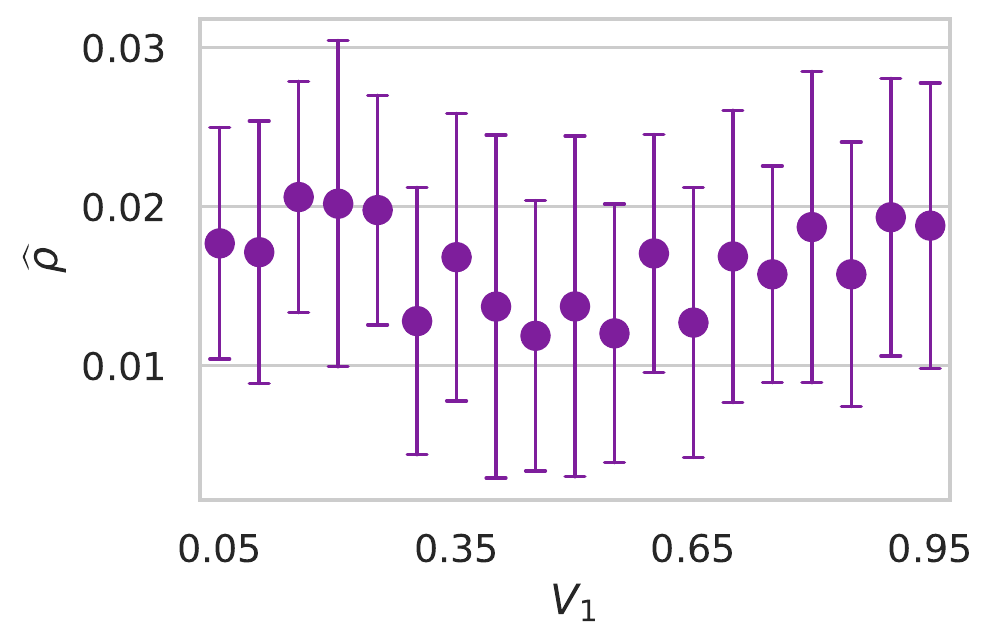}
}
\hbox{
\includegraphics[width=0.3\textwidth]{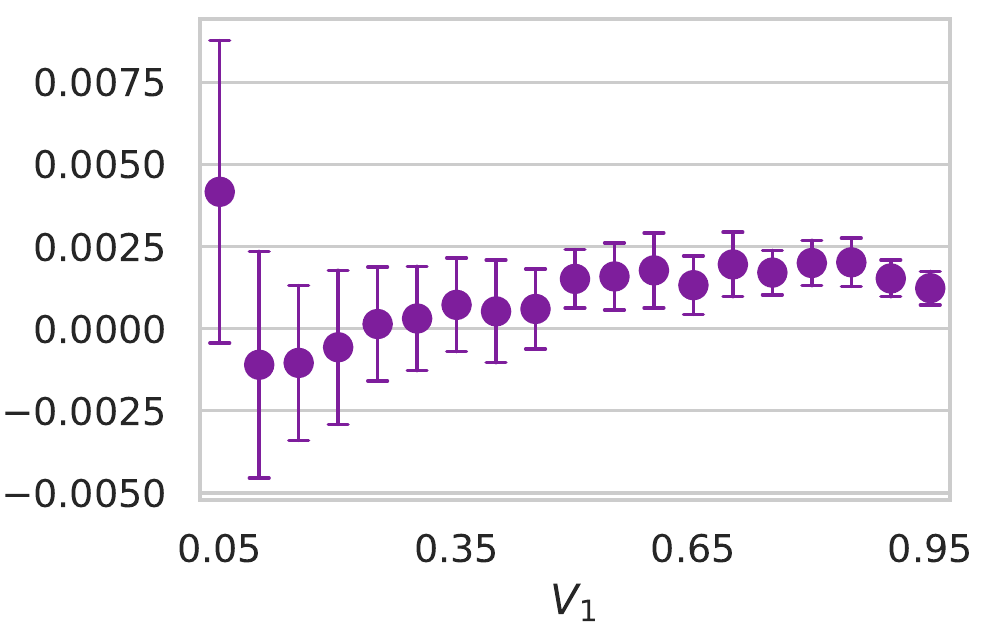}
\includegraphics[width=0.3\textwidth]{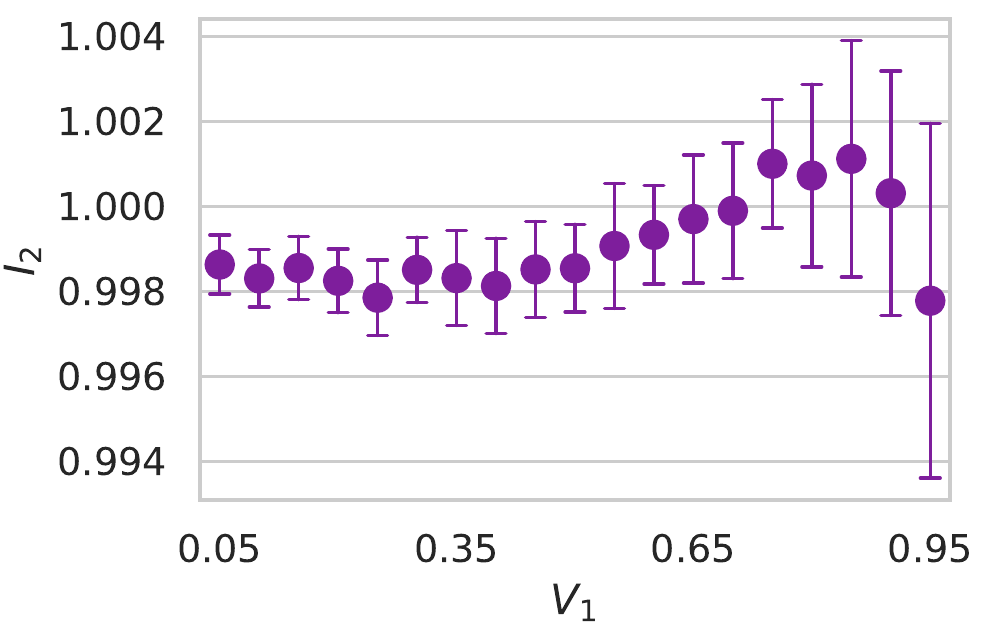}
\includegraphics[width=0.3\textwidth]{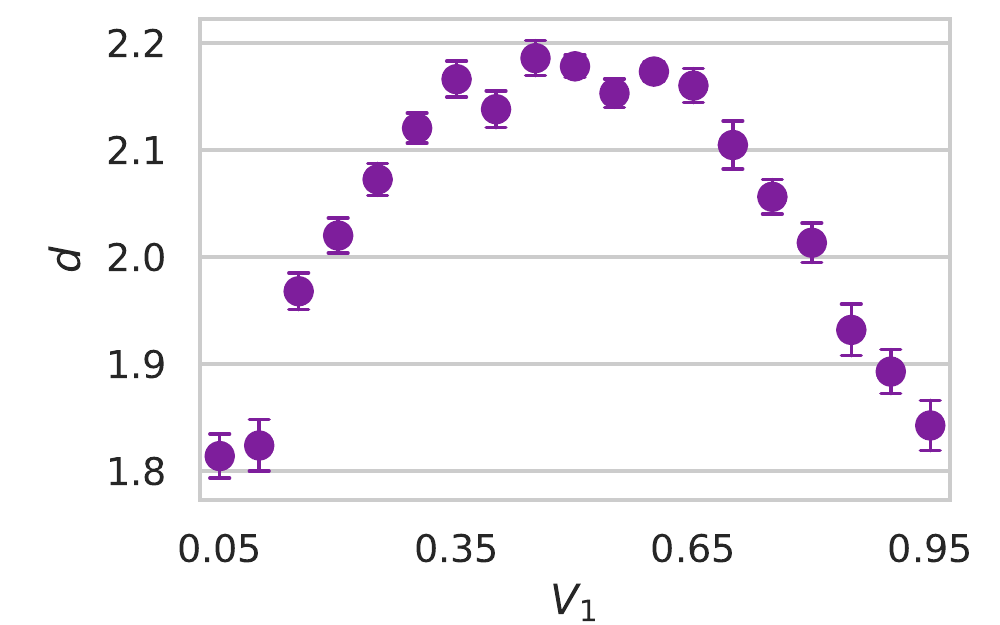}
}
\caption{\textbf{Model fitting results.} Model parameters estimated by the BIMM-2D for the 19 artificial datasets. Error bars indicate $\pm$ 1 standard deviation for 30 independent model fittings. The average values and standard deviations are listed in Table~\ref{tab:data_gen_vs_model_est} along with the parameters used when generating the data.}\label{fig:params_art}
\end{figure}
\begin{table}[!htp] %% batch size 50
\centering
\caption{Parameters used when generating the artificial data vs. parameters estimated using the BIMM-2D (mean and standard deviation of results for all datasets).}\label{tab:data_gen_vs_model_est}
%\begin{ruledtabular}
\begin{tabular}{llll}
\toprule
  \multicolumn{2}{l}{ Data generation} & \multicolumn{2}{l}{Model estimate} \\
 \hline\\
 $\sigma_b$ & 3 &  $\widehat{\sigma}_b$ & $3.04 \pm 0.04 $ \\
 $\sigma_n$ & 0.1 & $\widehat{\sigma}_n$ & $0.1009 \pm 0.0005 $\\
$\rho$ & 0 & $\widehat{\rho}$ &  $0.016 \pm 0.009 $\\
 $I_1$ & 0 & $\widehat{I}_1$ & $0.001 \pm 0.002 $\\
$I_2$ & 1 & $\widehat{I}_2$ & $0.999 \pm 0.002$ \\
\bottomrule \\
\end{tabular}
%\end{ruledtabular}
\end{table}
\begin{figure*}
\centering
    \subfloat[$L = -0.581$]{\includegraphics[width=0.64\textwidth]{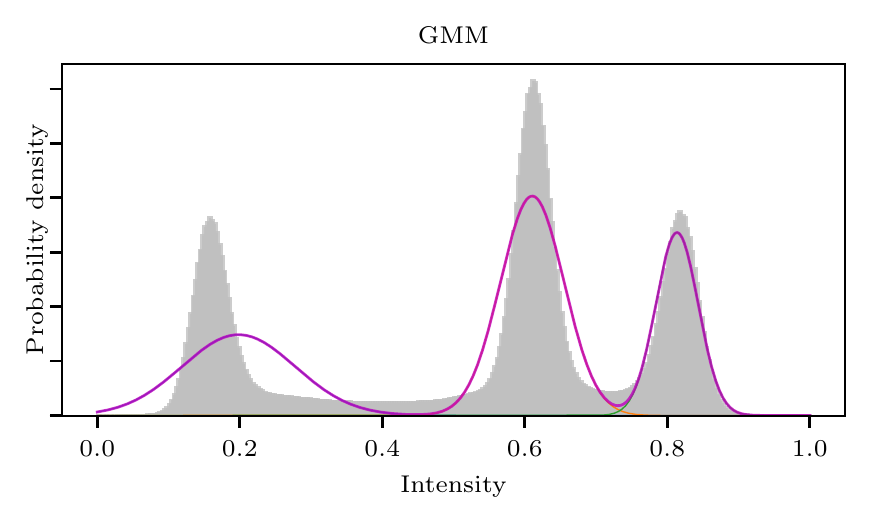}\label{fig:1Dhist_model_gmm_sup}}
    %add desired spacing between images, e. g. ~, \quad, \qquad, \hfill etc. 
      %(or a blank line to force the subfigure onto a new line)
      
    \subfloat[$L = -0.760$]{\includegraphics[width=0.64\textwidth]{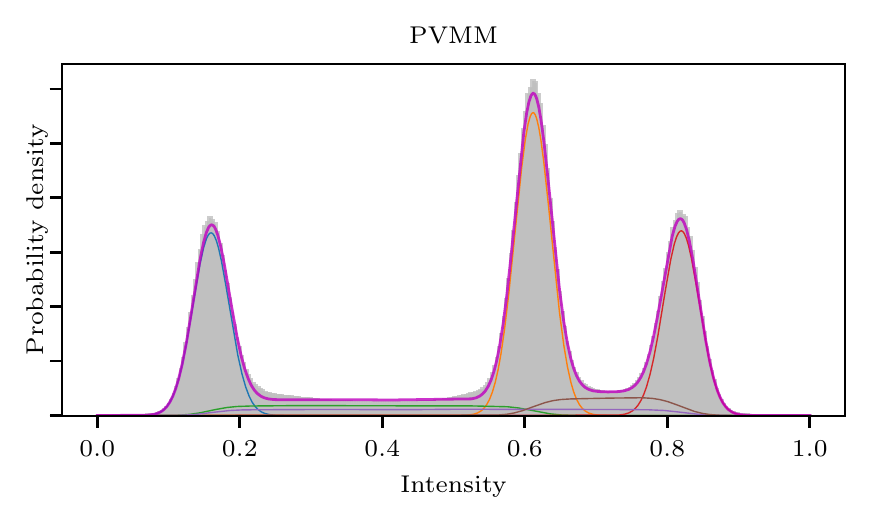}\label{fig:1Dhist_model_pvmm_sup}}
    
    \subfloat[$L = -0.762$]{\includegraphics[width=0.64\textwidth]{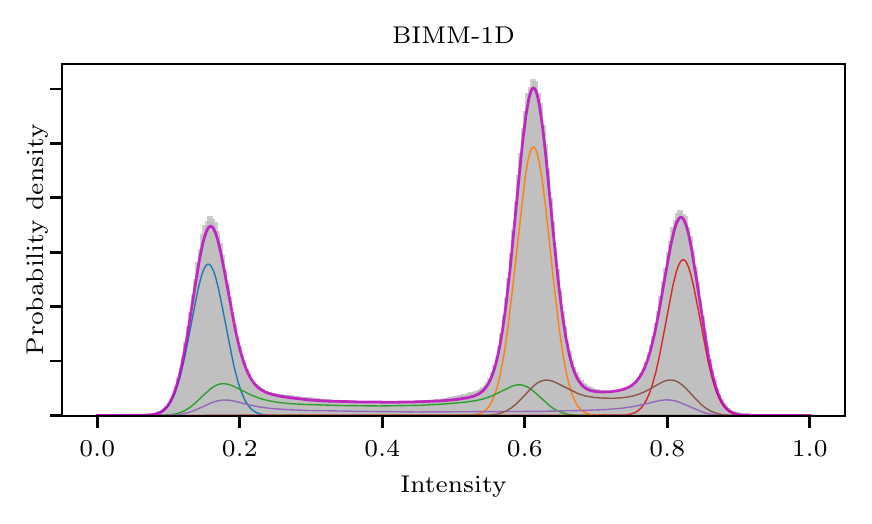}\label{fig:1Dhist_model_bimm1d_sup}}
    \caption{\textbf{Model fitting results.} Intensity histogram of the pristine fuel cell dataset (grey) with the PDF of fitted models (pink) and model components. A smaller value of the negative log-likelihood $L$ indicates a better model fit. }\label{fig:1Dhist_model_components}%(Eq.~\eqref{eq:L}) 
\end{figure*}